\documentclass[aps,prd,nofootinbib,showkeys,floatfix,twocolumn,bibnotes]{revtex4}
\usepackage{graphicx,amsmath,amssymb,hyperref,natbib,slashed}
\usepackage[dvipsnames]{xcolor}
\usepackage{xspace}
\usepackage{bbm} 

\usepackage[utf8]{inputenc}
\usepackage{placeins}

\newcommand{\eVdist}{\kern-0.06em}

\newcommand{\be}{\begin{equation}}
\newcommand{\ee}{\end{equation}}
\newcommand{\bea}{\begin{eqnarray}}
\newcommand{\eea}{\end{eqnarray}}

\begin{document}
\title{Strong constraints on self-interacting dark matter with light mediators}

\hfill \parbox{17.6cm}{\flushright DESY-16-226\vspace{1.5ex}}

\newcommand{\AddrCERN}{%
Theoretical Physics Department, CERN, Geneva, Switzerland
}
\newcommand{\AddrOslo}{%
Department of Physics, University of Oslo, Box 1048, N-0371 Oslo, Norway}
\newcommand{\AddrDESY}{%
Deutsches Elektronen-Synchrotron DESY,   Notkestra\ss e 85, D-22607 Hamburg, Germany}

 \author{Torsten Bringmann}
 \email{torsten.bringmann@fys.uio.no}
 \affiliation{\AddrOslo}
 
  \author{Felix Kahlhoefer}
 \email{felix.kahlhoefer@desy.de}
 \affiliation{\AddrDESY}
 
 \author{Kai Schmidt-Hoberg}
 \email{kai.schmidt-hoberg@desy.de}
 \affiliation{\AddrDESY}
 
  \author{Parampreet Walia}
 \email{p.s.walia@fys.uio.no}
 \affiliation{\AddrOslo~}

\begin{abstract}
Coupling dark matter to light new particles is an attractive way to combine thermal production with strong velocity-dependent self-interactions. Here we point out that in such models the dark matter annihilation rate is generically enhanced by the Sommerfeld effect, and we derive the resulting constraints from the Cosmic Microwave Background and other indirect detection probes. For the frequently studied case of s-wave annihilation these constraints exclude the entire parameter space where the self-interactions are large enough to address the small-scale problems of structure formation.
\end{abstract}

\maketitle


\paragraph*{Introduction.---}%
Although dark matter (DM) particles can only have very weak interactions with Standard Model (SM) states, it 
is an intriguing possibility that they experience much stronger self-interactions and thereby alter the behaviour 
of DM on astrophysical and cosmological scales in striking ways. In particular, self-interacting DM 
(SIDM) may offer an attractive solution to some of the long-standing small-scale structure problems 
encountered in the collisionless cold DM paradigm~\cite{deLaix:1995vi,Spergel:1999mh}. At the same time, a 
conclusive observation of collisionality of DM on astrophysical scales would have striking implications for the 
particle physics properties of DM. For these reasons, SIDM has been the subject of increasing interest in the 
last few years. 

In order to affect astrophysical observations, the DM self-scattering cross section typically has to be  
of order $\sigma / m_\chi \sim 1 \: \mathrm{cm^2\,g^{-1}}$ \cite{Buckley:2009in,Feng:2009hw,Feng:2009mn,Loeb:2010gj,Zavala:2012us,Vogelsberger:2012ku}. 
Such large cross sections can arise in essentially two different ways: either from new strong forces in the dark sector, 
similar to QCD, or from a more weakly coupled theory with a very light mediating particle. 
In the former case, large self-interactions are expected for all DM velocities, leading to strong bounds in particular from galaxy 
clusters~\citep{Markevitch:2003at,Randall:2007ph,Peter:2012jh,Rocha:2012jg,Kahlhoefer:2013dca,Harvey:2015hha,Kaplinghat:2015aga}. 
In contrast, in models with a very light mediator self-interactions become stronger at smaller DM velocities, so 
that large effects on small scales can be consistent with the stronger astrophysical constraints on larger 
scales~\cite{Ackerman:mha,Feng:2009mn,Buckley:2009in,Feng:2009hw,Feng:2009mn,Loeb:2010gj,Aarssen:2012fx,Tulin:2013teo,Kaplinghat:2015aga}.

Another attractive feature of SIDM with a very light mediator is that the DM abundance today can be 
explained by thermal production in the early Universe. In contrast to the standard WIMP 
scenario \cite{Gondolo:1990dk}, where DM annihilates directly into SM states (see~\cite{Chu:2016pew}), the 
DM particle $\chi$ annihilates into the mediator particles $\phi$, which subsequently decay into SM states: 
$\chi \chi \rightarrow \phi \phi \rightarrow \text{SM}$~\cite{Pospelov:2007mp}. 
This can naturally yield the observed relic abundance even if interactions with the SM are strongly suppressed.

From a particle physics perspective, the presence of very weakly coupled light mediators can be easily 
motivated, e.g.~if the stability of the DM particle arises from a charge under a new spontaneously 
broken $U(1)'$ gauge group. Nevertheless, the mediator should  couple to the SM at {\it some} 
level~\cite{Kaplinghat:2013yxa,DelNobile:2015uua} in order to {\it i)} bring the dark and visible sector  
into thermal equilibrium at early times, and {\it ii)} to decay sufficiently quickly to avoid overclosing the 
Universe. Such scenarios provide a both attractive and minimal realization 
of the SIDM idea, and have hence been one of the main avenues of model 
building~\cite{Kouvaris:2014uoa,Bernal:2015ova,Kainulainen:2015sva}.

The required interaction strength between mediators and the SM turns out to be rather small, and is 
therefore usually assumed to be essentially unconstrained. In this Letter we demonstrate that decays of the 
mediator into SM states are, instead, very strongly constrained for such models. The reason is that the large 
self-interaction cross sections are achieved via non-perturbative enhancements which at the same time also 
enhance the DM annihilation cross section, in particular for small DM 
velocities~\cite{Kamionkowski:2008gj,Zavala:2009mi,Feng:2010zp,Hisano:2011dc}. The subsequent decays of 
the mediators into SM states would then typically change the reionization history of the Universe, and thereby 
lead to significant distortions of the Cosmic Microwave Background (CMB), or be observed in indirect detection 
experiments today. Analogous constraints have been explored previously in the context of possible DM explanations of the cosmic-ray positron excess~\cite{ArkaniHamed:2008qn}, see 
e.g.~\cite{Bergstrom:2008ag,Mardon:2009rc, Galli:2009zc,Slatyer:2009yq,Zavala:2009mi,Feng:2010zp,Hannestad:2010zt,Finkbeiner:2010sm}. 
Here, we update these constraints specifically for SIDM, using the latest CMB and indirect detection data.

This Letter is organized as follows. We first revisit the required properties of the DM and mediator 
particles to reproduce the observed relic abundance and give rise to phenomenologically relevant DM 
self-interactions. We then discuss CMB and  indirect detection constraints, with special emphasis on  
strongly velocity-dependent DM annihilation rates. 
We illustrate their impact on the most popular class of models, in which the DM relic density is set by 
$s$-wave annihilation. Finally, we comment on how the resulting strong constraints may be relaxed.


\smallskip
\paragraph*{Self-interacting DM with light mediators.---}%
We consider a non-relativistic DM species $\chi$ that interacts via a light vector or scalar mediator $\phi$. The 
DM self-interactions that result from exchanging $\phi$ can be described by a Yukawa potential, 
which leads to a strong dependence of the self-interaction rate on the relative velocity $v$ of the scattering 
DM particles. The phenomenology of this scenario is  fully characterized by the two masses, $m_\chi$ and 
$m_\phi$, and the coupling strength $\alpha_\chi\equiv g^2/4\pi$. 

In the {\it Born limit} ($\alpha_\chi m_\chi \lesssim m_\phi$), the momentum transfer cross section $\sigma_T$ 
can be calculated perturbatively~\cite{Feng:2009hw}. For larger coupling strengths or DM masses, 
non-perturbative effects become important. In the following, we use the improved parameterization 
from \cite{Cyr-Racine:2015ihg} for the {\it classical limit} ($m_\chi v \gtrsim m_\phi$) and adopt the analytical 
expressions from \cite{Tulin:2013teo}, which have been obtained from approximating the Yukawa potential by a 
Hulth\'en potential, in the {\it intermediate (resonant) regime}. To estimate the effect of DM self-interactions on 
dwarf galaxies, we define $\langle\sigma_T\rangle_{30}$ as $\sigma_T$ averaged over a Maxwellian velocity 
distribution with a most probable velocity of $30\:\mathrm{km\,s^{-1}}$. To obtain observationally relevant 
effects, e.g.~to alleviate the cusp-core~\cite{deBlok:1997zlw,Oh:2010ea,Walker:2011zu} and 
too-big-to-fail \cite{BoylanKolchin:2011de,Papastergis:2014aba} problems,
we require $\langle\sigma_T\rangle_{30}/m_\chi\sim0.1\text{--}10\,\mathrm{cm^2\,g^{-1}}$~\cite{Zavala:2012us,Vogelsberger:2012ku}.

As motivated in the introduction, a second important constraint can be obtained under the assumption that the 
dark sector was in thermal 
equilibrium with the SM sector at early times and the DM relic abundance is set by thermal freeze-out. This 
assumption is well-motivated if the mediator couples also to SM states, but may need to be revisited if these 
interactions are very weak (see below). As the dominant DM annihilation channel is 
$\chi\chi\rightarrow\phi\phi$, we can effectively eliminate $\alpha$ as a free parameter by requiring that the 
relic density matches the observed value of $\Omega_\chi h^2=0.1188 \pm 0.0010$ 
\cite{Ade:2015xua}. 

The required value of $\alpha$ depends on the particle masses and on whether the annihilation proceeds via 
an $s$- or a $p$-wave process. The former implies a constant annihilation rate $(\sigma v)_0$ at the 
perturbative level, for \mbox{$v\ll1$}, while the latter implies $(\sigma v)_0\propto v^2$. In both 
cases, we can combine the requirement of sizeable self-interaction rates with the observed relic density for the 
three regimes mentioned above \cite{Tulin:2013teo}. As visualized later in Fig.~\ref{fig:swave}, this yields 
very roughly ($m_\chi\gtrsim100$\,GeV, $m_\phi\lesssim10$\,MeV) in the classical regime, 
($10\,\mathrm{GeV}\lesssim m_\chi\lesssim100$\,GeV, $1\,\mathrm{MeV}\lesssim m_\phi\lesssim1$\,GeV) 
in the resonant regime, and ($m_\chi\lesssim10$\,GeV, $m_\phi\lesssim10$\,MeV) in the Born limit.

\begin{figure}[t]
\includegraphics[width=\columnwidth]{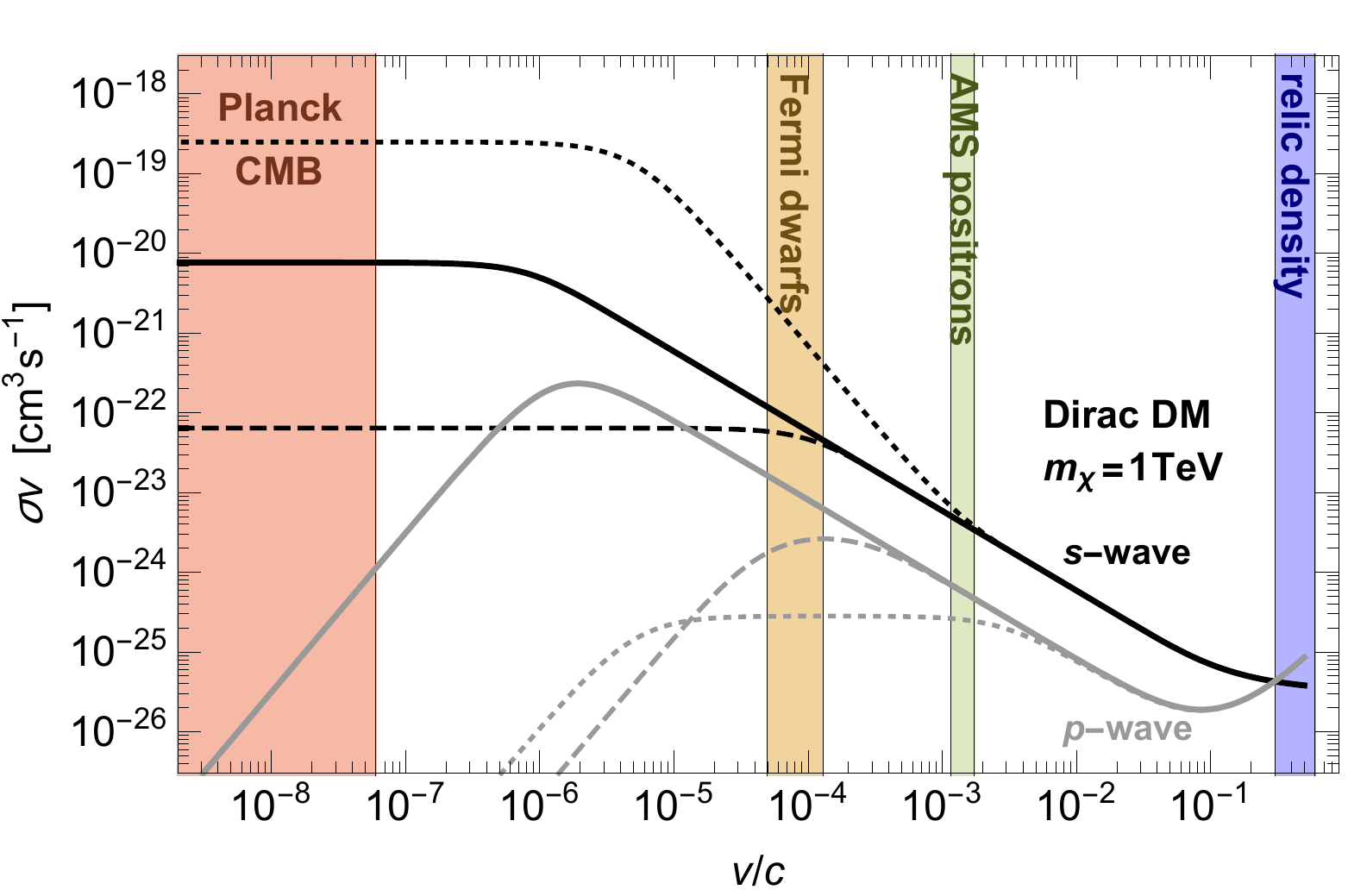}
\caption{Comparison of cross sections for $s$-wave and $p$-wave annihilation, as a function of the 
relative DM-DM velocity. The coupling $\alpha_\chi$ is fixed by the relic density requirement. 
Solid (dashed) curves correspond to $m_\chi=1$\,TeV and $m_\phi=1$\,MeV ($m_\phi=$100\,MeV),
while dotted lines show the case of $m_\phi$ tuned to 1.119\,GeV (1.066\,GeV) for resonant 
$s$-wave ($p$-wave) annihilation. 
In addition, the typical velocity ranges of different experimental probes are indicated.
\label{fig:spcomp}}
\end{figure}


\smallskip
\paragraph*{Sommerfeld enhancement.---}%
The Yukawa potential due to light mediator 
exchange does not only affect DM self-interactions, but it also modifies the wave-function of the annihilating DM 
pair \cite{Sommerfeld,ArkaniHamed:2008qn}. For small velocities, this can lead to significant non-perturbative 
corrections to the tree-level annihilation rate, \mbox{$\sigma v= S \times (\sigma v)_0$}, with the Sommerfeld 
enhancement factor $S$ given in~\cite{Cassel:2009wt,Iengo:2009ni,Slatyer:2009vg}.
For $\alpha_\chi m_\phi\ll m_\chi v^2$, 
the Yukawa potential becomes indistinguishable from a Coulomb potential and no strong resonances appear in  $S$.

This effect is usually taken into account for relic density calculations in SIDM models,
and we adopt here the results from \cite {Aarssen:2012fx, Tulin:2013teo}.
Quantitatively, the required value of $\alpha_\chi$ differs from the 
perturbative result only by an $\mathcal{O}(1)$ factor independent of $m_\phi$, because 
for most of the parameter space of interest we are in the Coulomb regime during chemical freeze-out.
We neglect the model-dependent effect of a second period of DM annihilation  {\it after kinetic decoupling} 
that can occur for Sommerfeld-enhanced DM annihilation \cite{Dent:2009bv,Zavala:2009mi,Feng:2010zp,vandenAarssen:2012ag}. While this may in principle
decrease $\Omega_\chi$ by up to three orders of magnitude if DM 
annihilation occurs very close to a resonance, it changes the calculation only at
the percent level off resonance \cite{vandenAarssen:2012ag}. Similarly, we neglect the effect of bound-state 
formation, which only becomes important close to the unitarity bound~\cite{vonHarling:2014kha} 
(for details see~\cite{Cirelli:2016rnw}).

As the Universe continues to cool down after DM freeze-out, the DM velocities decrease. 
The crucial observation for the purpose of this Letter is that the Sommerfeld enhancement at late times is
therefore much larger than during freeze-out. 
We illustrate this in Fig.~\ref{fig:spcomp}, where we show $s$-wave and 
$p$-wave annihilation cross sections as a function of the DM velocity for different mediator masses, 
with $\alpha_\chi$ being fixed by the relic density requirement. Away from any resonance, the enhancement 
scales like $1/v$ and $1/v^3$ for the $s$-wave and $p$-wave case, respectively, so that effectively the cross 
sections scale like $1/v$ in both cases. For the $p$-wave case, however, there is an offset compared to the 
thermal cross section due to the initial $v^2$ 
suppression. In both cases the saturation of the Sommerfeld enhancement occurs at about 
$v \sim m_\phi/2\hspace{0.25mm}m_\chi$, leading to a plateau for the $s$-wave and a maximum for the 
$p$-wave case. For masses tuned to a resonance (as shown for the dotted lines), the $s$-wave
enhancement grows like $1/v^2$ and saturates later. 

This figure clearly demonstrates that probes of DM annihilation at small velocities have the potential to 
seriously impact SIDM models with light mediators, in particular for $s$-wave annihilation. In the 
following, we will discuss the relevant observational constraints in turn and then quantify our general 
expectation by considering a popular concrete model realization that leads to $s$-wave dominated DM 
annihilation.

\smallskip
\paragraph*{CMB constraints.---}%
 Mediator particles decaying into SM particles at a rate larger than the 
Hubble rate can bring the model in conflict with the robust constraints on DM annihilation
from CMB observations \cite{Adams:1998nr,Chen:2003gz,Slatyer:2009yq,Galli:2009zc,Cline:2013fm,Liu:2016cnk}.  
At 95\%\,C.L. the most recent Planck data result in \cite{Ade:2015xua} 
\be
\label{eq:cmb_planck}
\frac{\langle\sigma v\rangle_\mathrm{rec}}{N_\chi} \lesssim 4 \times 10^{-25} \: \mathrm{cm^3 \, s}^{-1} \left(\frac{f_\mathrm{eff} }{0.1}\right)^{-1}\left(\frac{m_\chi}{100\,\mathrm{GeV}}\right)\,,
\ee
where $\langle\sigma v\rangle_\mathrm{rec}$ is the annihilation rate at recombination 
($z_\mathrm{rec}\sim 1100$) averaged over the distribution of DM velocities and
$N_\chi=1$ ($N_\chi=2$) for Majorana (Dirac) DM. The efficiency factor $f_\mathrm{eff}$ is 
related to the fraction of the released energy ending up in photons or electrons (see e.g.~\cite{Slatyer:2015jla}) 
with $f_\mathrm{eff} \gtrsim 0.1$  for any SM final state apart from neutrinos.

The typical DM velocity $v_\chi$ during recombination  depends on the 
temperature of kinetic decoupling $T_\mathrm{kd}$ via
$\langle v_\chi^2\rangle_\mathrm{rec} = \frac{1}{2} \langle v^2\rangle_\mathrm{rec} = 
3\, {(T_\mathrm{kd}}/{m_\chi)} \left({z_\mathrm{rec}}/{z_\mathrm{kd}}\right)^2 $
\cite{Bringmann:2009vf},
where $z_\mathrm{kd}$ is the redshift at kinetic decoupling.
For standard WIMPs, one expects 
$10\,\mathrm{MeV}\lesssim T_\mathrm{kd}\lesssim1\,\mathrm{GeV}$~\cite{Bringmann:2009vf}. In the 
presence of light mediators, however, kinetic decoupling can be significantly 
delayed \cite{vandenAarssen:2012ag}. Nevertheless, 
observations of the Lyman-$\alpha$ forest \cite{Croft:1997jf,Croft:2000hs} robustly exclude 
$T_\mathrm{kd}\lesssim100$\,eV (see Ref.~\cite{Vogelsberger:2015gpr, Bringmann:2016ilk} 
for a recent discussion). This translates into an {upper bound on the relative DM velocity at 
recombination} of 
\be
\label{eq:vrec}
 v_\mathrm{rec} \lesssim 2 \times10^{-7} \left(\frac{m_\chi}{100\,\mathrm{GeV}}  \right)^{-1/2}\,.
\ee
Such small velocities imply enormously enhanced DM annihilation rates. 

As an illustrative example, let us estimate the effect on the classical regime of DM self-scattering, assuming that
annihilation proceeds via an $s$-wave process. Since there are no resonances,  the 
Sommerfeld enhancement saturates for $v \lesssim v_\text{sat} \equiv m_\phi/2m_\chi$. 
For phenomenologically relevant values of $\langle\sigma_T\rangle_{30}$, 
this mass ratio is numerically 
larger than the minimal velocity in Eq.~(\ref{eq:vrec}). The annihilation 
rate relevant for CMB constraints thus becomes  {\it maximal}, 
\be
\label{eq:sigma_cmb}
\langle\sigma v\rangle_\mathrm{rec} \sim \langle\sigma v\rangle_\mathrm{cd} \frac{v^*}{v_\text{sat}} \sim \langle\sigma v\rangle_\mathrm{cd} \, v^* \, \frac{m_\chi}{m_\phi} \; ,
\ee 
where the annihilation rate at freeze-out is approximately 
$\langle\sigma v\rangle_\mathrm{cd} / N_\chi \sim 3 \times 10^{-26}\,\mathrm{cm^3 \, s^{-1}}$ in order to obtain 
the correct relic density and $v^*\sim0.1$ denotes the velocity below which $\sigma v\propto 1/v$.
Comparing Eq.~(\ref{eq:cmb_planck}) to Eq.~(\ref{eq:sigma_cmb}), we 
conclude that for $s$-wave annihilation in the classical scattering regime this class of models is ruled out unless 
either $m_\phi \gtrsim 1\:\text{GeV}$, in which case no sizeable DM self-interactions can be achieved, or 
$f_\mathrm{eff} \ll 0.1$.

\smallskip
\paragraph*{Other indirect detection constraints.---}%

In the Born and resonant regimes there are additional strong constraints from observations probing somewhat 
larger DM velocities, specifically from searches for present-day DM annihilation into light mediators 
$\phi$, which in turn 
decay into SM particles. For simplicity we will only consider light leptonic decay modes,
implying branching ratios $BR (\phi\to \ell\ell) < 1$ for  $m_\phi \gtrsim 2 \hspace{0.25mm} m_{\pi^0}$. 
Including further channels would lead to more stringent constraints. 

Dwarf galaxy observations with the Fermi gamma-ray space telescope provide one of the most robust
ways to constrain DM annihilation, and we implement them
using the likelihood functions provided by the 
Fermi-LAT collaboration which extend down to photon energies of $500\:\mathrm{MeV}$~\cite{Ackermann:2015zua}. 
To obtain the gamma-ray spectrum, we first calculate the distribution of photon energies from 
$\phi\to\ell^+\ell^-\gamma$, in the rest-frame of $\phi$, and then boost it to the DM frame as 
in \cite{Bergstrom:2008ag}. Kinematical data constrain the DM velocities to be much smaller than the 
$v_\chi\sim10^{-3}$ observed in our Galaxy; here, we adopt a relative velocity of $v=10^{-4}$, which 
independently of the assumed profile is a conservative choice \cite{Martinez:2013els}. 
We use the $J$-factors assuming a Burkert profile 
from~\cite{Ackermann:2013yva}.

Local DM annihilation to positrons are strongly constrained by the high-accuracy data of the AMS-02
experiment \cite{PhysRevLett.113.121101,PhysRevLett.113.121102}, with only moderate uncertainties related 
to the local normalizations of the DM
profile and radiation density \cite{Bergstrom:2013jra}. We take the bounds from \cite{Elor:2015bho} for 
one-step cascade annihilations with the intermediate state decaying to $e^+e^-$ and $\mu^+\mu^-$. These 
bounds extend down to DM masses of 10 GeV and, for $m_\phi \ll m_\chi$, are to good approximation 
independent of the mediator mass.

\begin{figure}[t!]
\includegraphics[width=0.9\columnwidth]{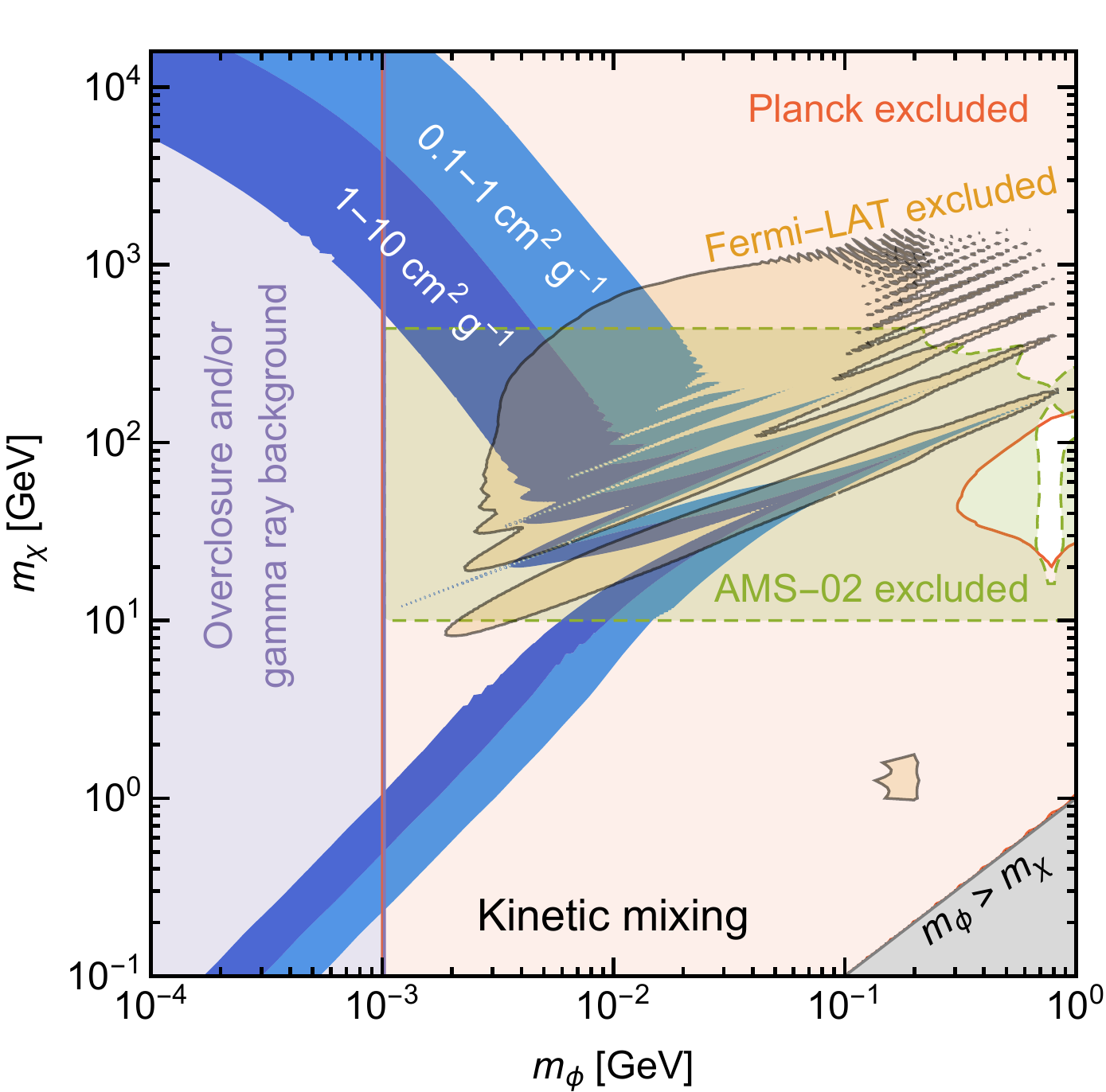}
\caption{Constraints at 95\%\,C.L.~on DM annihilating into vector mediators that 
kinematically mix with 
hypercharge as a function of the DM and mediator masses. The blue shaded region shows the combinations of 
DM mass $m_\chi$ and mediator mass $m_\phi$ that lead to a DM self-interaction cross section of  
$0.1\:\mathrm{cm^2\,g^{-1}}<\langle\sigma_T\rangle_{30}/m_\chi<10\:\mathrm{cm^2\,g^{-1}}$, which would 
visibly affect astrophysical observables at dwarf galaxy scale~\cite{Tulin:2013teo}. 
\label{fig:swave}}
\end{figure}

\smallskip
\paragraph*{Example model.---}%

Let us now apply the above constraints to an often discussed example of a model with 
$s$-wave DM annihilations~\cite{Pospelov:2008jd,ArkaniHamed:2008qn,Feng:2009mn,Tulin:2013teo,Kaplinghat:2013yxa}. 
Here, $\chi$ is a Dirac fermion that couples to a massive vector $\phi$.
The latter can obtain couplings to SM particles via kinetic mixing with the hypercharge field strength 
$B^{\mu\nu}$ or via mass mixing with the 
$Z$~\cite{Feldman:2006wd,Frandsen:2011cg,Chu:2016pew}:
\begin{equation}
\label{eq:Ls}
\mathcal{L} \supset - g^\mathrm{V}_\chi \phi^\mu \bar{\chi}\gamma_\mu \chi - \frac{1}{2} \sin \epsilon\, B_{\mu\nu} \phi^{\mu\nu} -\delta m^2 \phi^\mu Z_\mu \; .
\end{equation}
We first focus on the case of negligible mass mixing, $\delta m\ll m_\phi$. For $m_\phi \ll m_Z$ the couplings of 
the mediator are then largely photon-like, so this situation is very similar to kinetic mixing with 
electromagnetism~\cite{Foot:2004pa,Redondo:2008ec}. The dominant 
decay mode for 
$m_\phi\lesssim1$\,MeV is then $\phi \rightarrow 3\gamma$, which is so small that $\phi$ would 
effectively be stable; in such a scenario the mediator would either overclose the Universe or, close to the 
threshold, produce gamma rays in excess of the extragalactic gamma-ray background \cite{Redondo:2008ec}. 
Heavier mediators decay into both leptons and hadrons, and we adopt 
$BR (\phi\to \ell\ell)$ from~\cite{Curtin:2014cca}.

For each combination of DM and mediator mass in this model, we calculate the Sommerfeld enhancement 
factor using the conservative upper bound on $v_\mathrm{rec}$ from Eq.~(\ref{eq:vrec}). By comparing the 
result to Eq.~(\ref{eq:cmb_planck}), we can determine the parameter region excluded by CMB constraints. To 
calculate the appropriate value of $f_\mathrm{eff}$ as a function of $m_\chi$ and $m_\phi$, we multiply the 
different decay modes with the efficiency factors from~\cite{Slatyer:2015jla}. 
Our results are shown in Fig.~\ref{fig:swave}, where we also show the Fermi and AMS-02 bounds discussed
above. We observe that the CMB constraints, and partially also the other indirect detection constraints, 
exclude all combinations of $m_\chi$ and $m_\phi$ that lead to interesting self-interaction cross sections
(note that for sufficiently light DM the model is already excluded by these constraints without Sommerfeld enhancement). 

We emphasize that very close to a resonance both the preferred SIDM region and the various constraints may 
be modified by the impact of a potential second period of DM annihilation on the relic density calculation (see 
above). For late kinetic decoupling the resulting modifications will be small, but we expect even larger 
effects not to change our results qualitatively.

\smallskip
\paragraph*{Discussion.---}%

The bounds shown in Fig.~\ref{fig:swave} have been obtained under very conservative assumptions and are 
expected to apply in a similar way to other DM models
with a light mediator, e.g.~scalar and vector DM, for which annihilation into mediator pairs 
generically proceeds via $s$-wave. The CMB constraints, 
in particular,  are very robust because we probe DM annihilation in a kinematical 
situation where the Sommerfeld enhancement is typically already saturated,
so that the redshift dependence of the energy injection rate is the same as for 
standard $s$-wave annihilation. Even for parameter combinations 
where this is not the case, our constraints are extremely conservative because we evaluate  
$\sigma v$ no later than at recombination, and for larger values of $v_\mathrm{rec}$ than expected in 
a realistic treatment of kinetic decoupling. Nevertheless, our analysis does rely on a number of 
assumptions, which we will now review in detail.

For our calculations so far, there was no need to specify the kinetic mixing parameter 
$\epsilon$, as long as mixing is sufficiently large that the mediator decays in time to affect the reionisation 
history. Nevertheless, we have assumed implicitly that $\epsilon$ is large enough to thermalise 
the visible sector and the dark sector before freeze-out. Depending on the DM mass, the required 
value of $\epsilon$ 
for this to happen is of order $10^{-7}\text{--}10^{-5}$~\cite{Chu:2011be}. However, DM direct detection 
experiments (as well as astrophysical constraints for $m_\phi \lesssim 1\:\text{MeV}$~\cite{Redondo:2013lna}) 
typically require much smaller values of $\epsilon$~\cite{Kaplinghat:2013yxa}. The conclusion is that a different 
mechanism must be responsible for bringing the visible and the dark sector into thermal contact.

The simplest possibility would be a thermal contact at higher temperatures, via a different portal. 
After this interaction ceases to be effective,  the temperatures of both sectors would then evolve 
independently, depending on the 
number of degrees of freedom in each sector. For sizeable $\alpha_\chi$ the DM relic abundance will still be 
determined by dark sector freeze out, but at a different temperature. For reasonable temperature ratios, as we 
discuss in detail in appendix~\ref{app:thermal}, such a situation does not lead to qualitatively different results 
compared to the case where the two sectors have the same temperature. For the case where the two sectors 
never reach thermal equilibrium and the DM relic abundance is for example set via the freeze-in mechanism, 
we refer to~\cite{Bernal:2015ova}.

A second important assumption is that the DM annihilation to mediator pairs proceeds via an $s$-wave 
process. While the Sommerfeld 
enhancement can be significant also in the $p$-wave case (see Fig.~\ref{fig:spcomp}), the resulting cross 
sections are significantly smaller and indirect detection gives no relevant constraints. CMB constraints are also 
evaded for most of the parameter space, because for $v \lesssim v_\text{sat}$ the cross section again 
decreases like $v^2$ and therefore becomes unobservably small at recombination. Only if the ratio 
$m_\chi / m_\phi$ is very large and close to a resonance, effects may be observable -- in particular with 
stage 4 CMB experiments. Detailed predictions depend however on $v_\text{rec}$ and hence on 
$T_\text{kd}$. We note that $p$-wave annihilation requires scalar rather than vector mediators, which  
is strongly constrained from independent model building considerations, in particular the combination 
of constraints from direct detection experiments and primordial  nucleosynthesis~\cite{Kaplinghat:2013yxa}. 

Finally, our conclusions can be modified if the mediator decays in a different way than via kinetic 
mixing. As a specific example, we discuss the case of mass mixing in appendix~\ref{app:mass}.
In this case the mediator obtains a significant coupling to neutrinos, which alleviates constraints from both DM 
annihilation and the mediator lifetime, but in principle offers exciting prospects for indirect 
detection \cite{Aarssen:2012fx}: DM annihilation into a pair of mediators followed by the decay 
$\phi\to\bar\nu\nu$ would result in a 
characteristic spectral feature~\cite{Garcia-Cely:2016pse}. 
While currently unconstrained for the models considered here, such a signal is in reach for IceCube 
observations of the Galactic halo \cite{IceCube:2011ae,Dasgupta:2012bd,Aartsen:2015xej, Aisati:2015vma}.

In general, however, the constraints derived above are so strong that they can even be applied to models 
where mediator decays into leptons are sub-dominant. As a result, large self-interactions are excluded also 
for the case of mass mixing, as long as $m_\phi > 2 \hspace{0.25mm} m_e$. Even weaker constraints 
could in principle be obtained if the mediators couple to another very light state in the dark sector, such as 
sterile neutrinos. Such models are particularly interesting because they can significantly delay kinetic 
decoupling and thus provide a solution also to the missing satellite problem~\cite{Aarssen:2012fx,
Bringmann:2013vra,Dasgupta:2013zpn,Chu:2014lja,Ko:2014bka,Binder:2016pnr,Bringmann:2016ilk}.

\smallskip
\paragraph*{Conclusions.---}%
Models of DM with velocity-dependent self-interactions have recently received a great deal of attention for their 
potential to produce a number of interesting effects on astrophysical scales. 
We have shown in this Letter that these models face very strong constraints from the CMB and DM indirect 
detection. In the most natural realization of this scenario with a light vector mediator with kinetic mixing, these 
constraints rule out the entire parameter space where the self-scattering cross section can be relevant for 
astrophysical systems. These bounds remain highly relevant for a number of generalizations of the scenario, 
such as a different dark sector temperature and different mediator branching ratios. Clearly, future efforts to 
develop particle physics models for SIDM need to address these issues in order to arrive at models that provide 
a picture consistent with all observations in cosmology, astrophysics and particle physics.

\bigskip
 \paragraph*{Acknowledgements.---}%
 We thank Camilo Garcia-Cely, Michael Gustafsson, Julian Heeck, Andrzej Hryczuk, Joerg Jaeckel, 
 Manoj Kaplinghat, Andreas Ringwald, Marco Taoso, Sebastian Wild and Bryan Zaldivar for 
 enlightening discussions.
This work is supported by the German Science Foundation (DFG) under the
Collaborative Research Center (SFB) 676 Particles, Strings and the Early Universe as well as the
ERC Starting Grant `NewAve' (638528). P.~W.~is partially supported by the University of Oslo 
through the Strategic Dark Matter Initiative (SDI).


\appendix

\section*{Appendix}

In the main text we have considered constraints arising from the $s$-wave annihilations of SIDM, with a special focus on the case that DM self-interactions proceed via a light vector mediator. Our results were derived under the assumption that the vector mediator obtains couplings to SM states from kinetic mixing with hypercharge and that the dark and visible sectors are in thermal equilibrium at DM freeze-out. In this appendix we review these assumptions. In Sec.~\ref{app:thermal} we discuss the possibility of an independent temperature evolution of the two sectors and then consider the case of mass mixing between the mediator and the $Z$ boson in Sec.~\ref{app:mass}.

\section{Modifying the dark sector temperature}
\label{app:thermal}

If the couplings of the new mediator to the SM sector become very small, the dark and visible sectors will no 
longer be able to thermalize with each other in the early Universe. In principle, the temperature of the dark 
sector can then be almost arbitrary 
and could for example be set by the details of reheating. Nevertheless, it makes sense to assume that some 
other mechanism (e.g.\ another \emph{heavy} mediator) brings the two sectors into thermal equilibrium at high 
temperatures. Once  this mechanism becomes inefficient, the temperatures of 
the two sectors will evolve independently and will in general differ due to the different number of degrees of 
freedom in the two sectors.

Denoting the ratio of effective relativistic degrees of freedom in the dark sector and the visible sector by 
\mbox{$\eta \equiv g_{\ast,\text{dark}} / g_{\ast,\text{vis}}$}, the temperature ratio between the two sectors, 
\mbox{$\xi \equiv T_\text{dark} / T_\text{vis}$}, is given by 
\mbox{$\xi(T) = \left[\eta(T_\text{dec})/\eta(T)\right]^{1/3}$} as long as entropy is conserved separately
in the two sectors. Here, $T$ denotes the standard photon 
temperature, and $T_\text{dec}$ its value right after decoupling.
Since the number of effective degrees of freedom in the visible sector typically decreases more rapidly with 
decreasing temperature, in particular if one does not want to proliferate the number of new states, 
this places a lower bound on the temperature ratio. 
If the mediator and DM particles are the only new states in the dark sector right after decoupling, for example,
$g_{\ast,\text{dark}}$ does not  change at all and this lower bound is given by 
$\xi (T)\gtrsim 0.34\, [g_\text{*,vis}(T_\text{dec})/100]^{-1/3}$ when conservatively assuming
that the thermal freeze-out of DM happens as late as at $T_\text{cd}\lesssim1$\,MeV \cite{Bringmann:2016ilk} 
(in reality, we have $T_\text{cd}\sim m_\chi/25$, leading to considerably larger lower bounds on $\xi$ for
DM heavier than about 100\,MeV).

\begin{figure}[t]
\includegraphics[width=0.85\columnwidth]{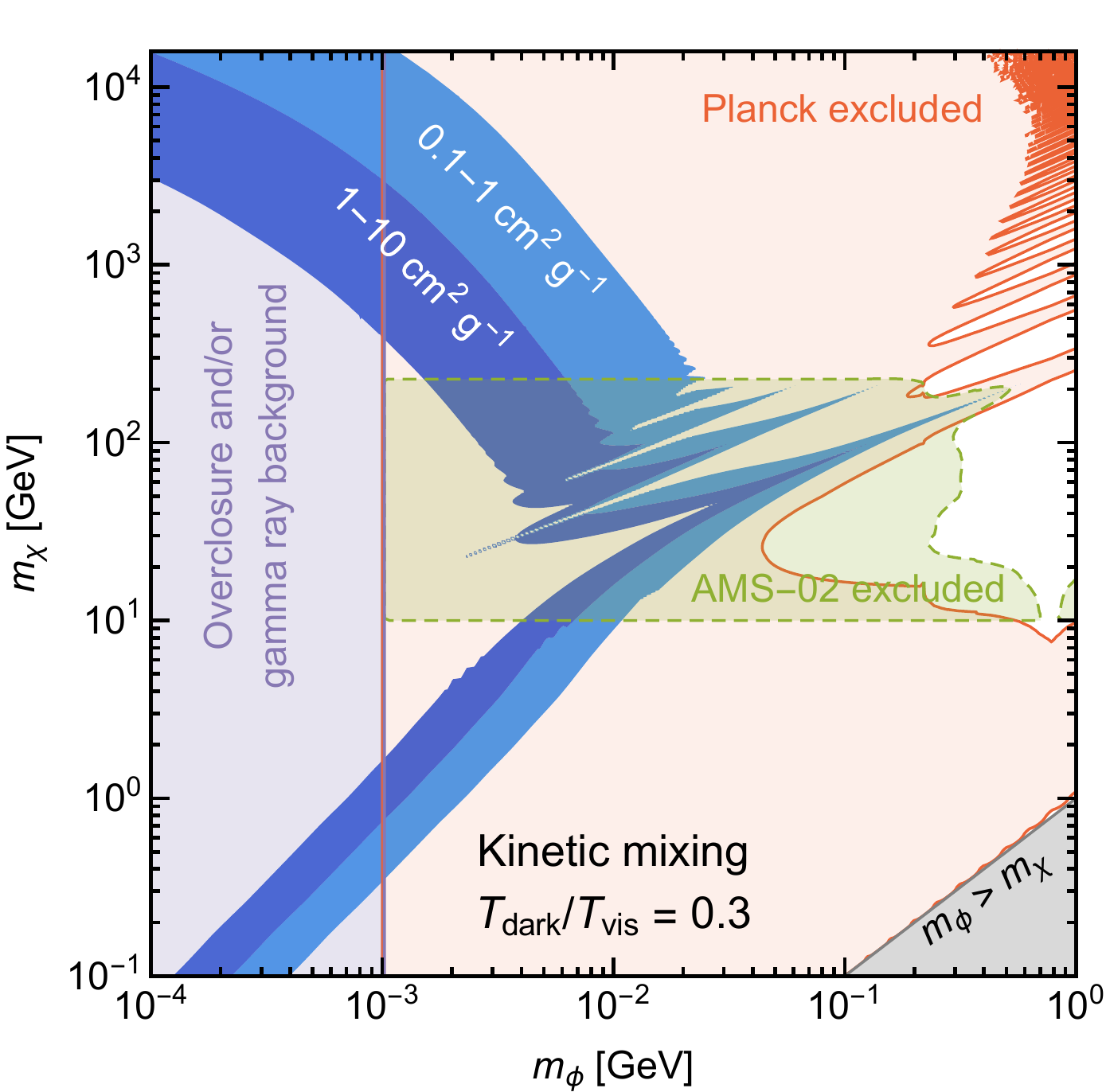}
\vspace*{-0.2cm}
\caption{Constraints on DM particles annihilating into vector mediators that kinematically mix with 
hypercharge, assuming a temperature ratio of $\xi=0.3$ between dark and visible sector during 
the thermal freeze-out of DM.
\label{fig:xi}}
\vspace*{-0.3cm}
\end{figure}

This change in temperature changes the value of $\alpha_\chi$ that gives the correct relic abundance. To first 
approximation, and neglecting additional mass-dependent corrections, one can replace 
$\alpha_\chi \rightarrow \sqrt{\xi}\alpha_\chi$~\cite{Feng:2008mu,Kaplinghat:2015gha,Bringmann:2016ilk}. 
Since the late time 
annihilation and the self scattering cross sections scale differently with $\alpha_\chi$, the constraints can be 
somewhat alleviated, as shown in Fig.~\ref{fig:xi} for the rather extreme case of $\xi=0.3$. 

While it is conceivable that the CMB constraints can be evaded for even smaller values of $\xi$, 
such extreme temperature ratios are difficult to reconcile with the assumption that
the dark and visible sector were in full thermal equilibrium at {\it some} high temperature. 
Moreover, if $\xi$ is very different from unity the validity of the simple rescaling 
$\alpha_\chi \rightarrow \sqrt{\xi}\alpha_\chi$ breaks down, so the relic density must be calculated 
from a numerical solution of the full Boltzmann equations instead 
(see, e.g., Ref.~\cite{Bringmann:2016ilk}).

\section{Mass mixing}
\label{app:mass}

Let us now consider the case that the decay modes of the mediator are dominantly set by its mass mixing with 
the SM $Z$ boson, which requires $\delta m / m_Z \gg \epsilon$. The most important change in this case is  
that the mediator obtains sizeable couplings to neutrinos. As a result, the overclosure and gamma-ray 
constraints for $m_\phi < 2 \hspace{0.25mm} m_e$ can be evaded, as it is easily possible for such light 
mediators to decay into neutrinos before primordial nucleosynthesis. The mixing required for such decays is still 
sufficiently small that it is essentially unconstrained by neutrino beamline experiments like 
LSND~\cite{deNiverville:2011it}.

We calculate the branching ratios for the case of mass mixing by analytically rescaling the branching ratios for 
kinetic mixing, see~\cite{Frandsen:2011cg}. We find that even for \mbox{$m_\chi > 2 \hspace{0.25mm} m_e$} 
the mediator decays dominantly invisibly up to the point where hadronic resonances become important. As a 
result, all our constraints are significantly weakened (see 
\begin{figure}[t]
\includegraphics[width=0.85\columnwidth]{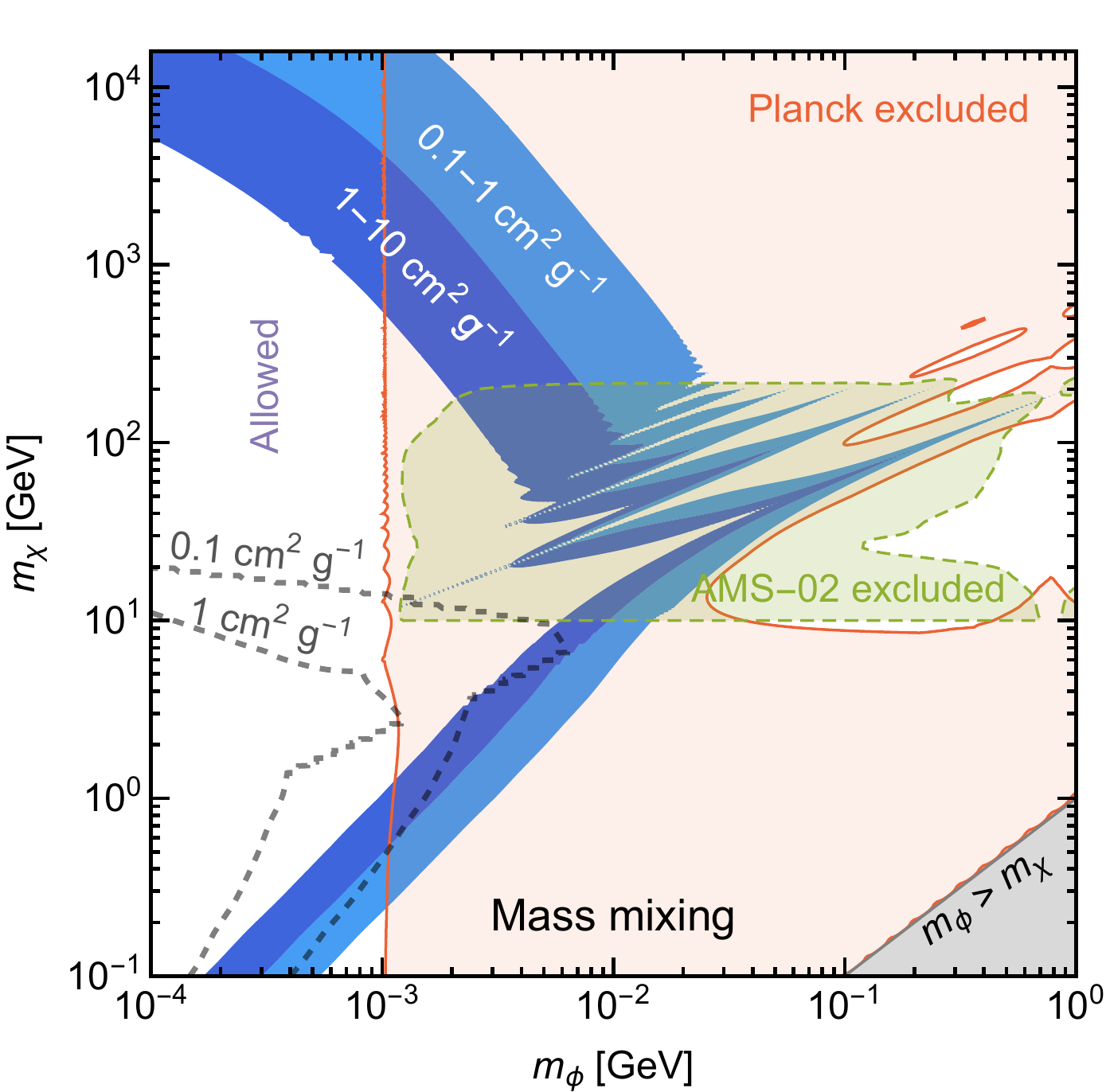}
\vspace*{-0.2cm}
\caption{Constraints on DM particles annihilating into vector mediators that couple to the SM via mass mixing 
with the $Z$ boson. The black dashed lines indicate constant values of $\langle \sigma_T \rangle_{1000}$ 
corresponding to astrophysically relevant self-interaction cross sections on cluster scales.
\label{fig:mass}}
\vspace*{-0.3cm}
\end{figure}
\FloatBarrier
\noindent Fig.~\ref{fig:mass}). Nevertheless, the parameter 
region required 
to obtain sizeable self-interaction cross sections is still solidly excluded by CMB constraints, as 
long as \mbox{$m_\phi > 2 \hspace{0.25mm} m_e$}. 
For smaller mediator masses and 
$m_\chi > 1 \:\text{TeV}$ it is possible to evade the experimental constraints discussed above and still
have self-interactions rates of a phenomenologically relevant strength.

For $m_\phi \lesssim 1 \:\mathrm{MeV}$ and $m_\chi \lesssim 10\:\mathrm{GeV}$ bounds on the 
self-interaction cross section at cluster scales become 
relevant~\cite{Ackerman:mha,Feng:2009mn,Buckley:2009in,Loeb:2010gj,Aarssen:2012fx,Tulin:2013teo}. While 
there is no general consensus regarding the precise value of these bounds (which may even depend on the 
details of the underlying particle physics), cross sections above $1\:\mathrm{cm^2\,g^{-1}}$ are likely excluded, 
while cross sections below $0.1\:\mathrm{cm^2\,g^{-1}}$ are likely consistent with all observations~\cite{Markevitch:2003at,Randall:2007ph,Peter:2012jh,Rocha:2012jg,Kahlhoefer:2013dca,Harvey:2015hha}. We indicate 
these two values of $\langle \sigma_T \rangle_{1000}$ with dashed black lines in Fig.~\ref{fig:mass} to give an 
estimate for the strength of these constraints. One can see that cluster constraints are much more easily 
evaded in the classical regime than in the Born regime, where the velocity dependence of the self-interaction 
cross section is rather weak. These considerations apply of course in the same way also to 
Figs.~2 and \ref{fig:xi}, where the corresponding regions of parameter space are already 
excluded by other observations and hence the cluster constraints are not shown explicitly.


\begin{thebibliography}{85}%
\makeatletter
\providecommand \@ifxundefined [1]{%
 \@ifx{#1\undefined}
}%
\providecommand \@ifnum [1]{%
 \ifnum #1\expandafter \@firstoftwo
 \else \expandafter \@secondoftwo
 \fi
}%
\providecommand \@ifx [1]{%
 \ifx #1\expandafter \@firstoftwo
 \else \expandafter \@secondoftwo
 \fi
}%
\providecommand \natexlab [1]{#1}%
\providecommand \enquote  [1]{``#1''}%
\providecommand \bibnamefont  [1]{#1}%
\providecommand \bibfnamefont [1]{#1}%
\providecommand \citenamefont [1]{#1}%
\providecommand \href@noop [0]{\@secondoftwo}%
\providecommand \href [0]{\begingroup \@sanitize@url \@href}%
\providecommand \@href[1]{\@@startlink{#1}\@@href}%
\providecommand \@@href[1]{\endgroup#1\@@endlink}%
\providecommand \@sanitize@url [0]{\catcode `\\12\catcode `\$12\catcode
  `\&12\catcode `\#12\catcode `\^12\catcode `\_12\catcode `\%12\relax}%
\providecommand \@@startlink[1]{}%
\providecommand \@@endlink[0]{}%
\providecommand \url  [0]{\begingroup\@sanitize@url \@url }%
\providecommand \@url [1]{\endgroup\@href {#1}{\urlprefix }}%
\providecommand \urlprefix  [0]{URL }%
\providecommand \Eprint [0]{\href }%
\providecommand \doibase [0]{http://dx.doi.org/}%
\providecommand \selectlanguage [0]{\@gobble}%
\providecommand \bibinfo  [0]{\@secondoftwo}%
\providecommand \bibfield  [0]{\@secondoftwo}%
\providecommand \translation [1]{[#1]}%
\providecommand \BibitemOpen [0]{}%
\providecommand \bibitemStop [0]{}%
\providecommand \bibitemNoStop [0]{.\EOS\space}%
\providecommand \EOS [0]{\spacefactor3000\relax}%
\providecommand \BibitemShut  [1]{\csname bibitem#1\endcsname}%
\let\auto@bib@innerbib\@empty
\bibitem [{\citenamefont {de~Laix}\ \emph {et~al.}(1995)\citenamefont
  {de~Laix}, \citenamefont {Scherrer},\ and\ \citenamefont
  {Schaefer}}]{deLaix:1995vi}%
  \BibitemOpen
  \bibfield  {author} {\bibinfo {author} {\bibfnamefont {A.~A.}\ \bibnamefont
  {de~Laix}}, \bibinfo {author} {\bibfnamefont {R.~J.}\ \bibnamefont
  {Scherrer}}, \ and\ \bibinfo {author} {\bibfnamefont {R.~K.}\ \bibnamefont
  {Schaefer}},\ }\href {\doibase 10.1086/176322} {\bibfield  {journal}
  {\bibinfo  {journal} {Astrophys. J.}\ }\textbf {\bibinfo {volume} {452}},\
  \bibinfo {pages} {495} (\bibinfo {year} {1995})},\ \Eprint
  {http://arxiv.org/abs/astro-ph/9502087} {arXiv:astro-ph/9502087 [astro-ph]}
  \BibitemShut {NoStop}%
\bibitem [{\citenamefont {Spergel}\ and\ \citenamefont
  {Steinhardt}(2000)}]{Spergel:1999mh}%
  \BibitemOpen
  \bibfield  {author} {\bibinfo {author} {\bibfnamefont {D.~N.}\ \bibnamefont
  {Spergel}}\ and\ \bibinfo {author} {\bibfnamefont {P.~J.}\ \bibnamefont
  {Steinhardt}},\ }\href {\doibase 10.1103/PhysRevLett.84.3760} {\bibfield
  {journal} {\bibinfo  {journal} {Phys. Rev. Lett.}\ }\textbf {\bibinfo
  {volume} {84}},\ \bibinfo {pages} {3760} (\bibinfo {year} {2000})},\ \Eprint
  {http://arxiv.org/abs/astro-ph/9909386} {arXiv:astro-ph/9909386 [astro-ph]}
  \BibitemShut {NoStop}%
\bibitem [{\citenamefont {Buckley}\ and\ \citenamefont
  {Fox}(2010)}]{Buckley:2009in}%
  \BibitemOpen
  \bibfield  {author} {\bibinfo {author} {\bibfnamefont {M.~R.}\ \bibnamefont
  {Buckley}}\ and\ \bibinfo {author} {\bibfnamefont {P.~J.}\ \bibnamefont
  {Fox}},\ }\href {\doibase 10.1103/PhysRevD.81.083522} {\bibfield  {journal}
  {\bibinfo  {journal} {Phys. Rev.}\ }\textbf {\bibinfo {volume} {D81}},\
  \bibinfo {pages} {083522} (\bibinfo {year} {2010})},\ \Eprint
  {http://arxiv.org/abs/0911.3898} {arXiv:0911.3898 [hep-ph]} \BibitemShut
  {NoStop}%
\bibitem [{\citenamefont {Feng}\ \emph
  {et~al.}(2010{\natexlab{a}})\citenamefont {Feng}, \citenamefont
  {Kaplinghat},\ and\ \citenamefont {Yu}}]{Feng:2009hw}%
  \BibitemOpen
  \bibfield  {author} {\bibinfo {author} {\bibfnamefont {J.~L.}\ \bibnamefont
  {Feng}}, \bibinfo {author} {\bibfnamefont {M.}~\bibnamefont {Kaplinghat}}, \
  and\ \bibinfo {author} {\bibfnamefont {H.-B.}\ \bibnamefont {Yu}},\ }\href
  {\doibase 10.1103/PhysRevLett.104.151301} {\bibfield  {journal} {\bibinfo
  {journal} {Phys. Rev. Lett.}\ }\textbf {\bibinfo {volume} {104}},\ \bibinfo
  {pages} {151301} (\bibinfo {year} {2010}{\natexlab{a}})},\ \Eprint
  {http://arxiv.org/abs/0911.0422} {arXiv:0911.0422 [hep-ph]} \BibitemShut
  {NoStop}%
\bibitem [{\citenamefont {Feng}\ \emph {et~al.}(2009)\citenamefont {Feng},
  \citenamefont {Kaplinghat}, \citenamefont {Tu},\ and\ \citenamefont
  {Yu}}]{Feng:2009mn}%
  \BibitemOpen
  \bibfield  {author} {\bibinfo {author} {\bibfnamefont {J.~L.}\ \bibnamefont
  {Feng}}, \bibinfo {author} {\bibfnamefont {M.}~\bibnamefont {Kaplinghat}},
  \bibinfo {author} {\bibfnamefont {H.}~\bibnamefont {Tu}}, \ and\ \bibinfo
  {author} {\bibfnamefont {H.-B.}\ \bibnamefont {Yu}},\ }\href {\doibase
  10.1088/1475-7516/2009/07/004} {\bibfield  {journal} {\bibinfo  {journal}
  {JCAP}\ }\textbf {\bibinfo {volume} {0907}},\ \bibinfo {pages} {004}
  (\bibinfo {year} {2009})},\ \Eprint {http://arxiv.org/abs/0905.3039}
  {arXiv:0905.3039 [hep-ph]} \BibitemShut {NoStop}%
\bibitem [{\citenamefont {Loeb}\ and\ \citenamefont
  {Weiner}(2011)}]{Loeb:2010gj}%
  \BibitemOpen
  \bibfield  {author} {\bibinfo {author} {\bibfnamefont {A.}~\bibnamefont
  {Loeb}}\ and\ \bibinfo {author} {\bibfnamefont {N.}~\bibnamefont {Weiner}},\
  }\href {\doibase 10.1103/PhysRevLett.106.171302} {\bibfield  {journal}
  {\bibinfo  {journal} {Phys. Rev. Lett.}\ }\textbf {\bibinfo {volume} {106}},\
  \bibinfo {pages} {171302} (\bibinfo {year} {2011})},\ \Eprint
  {http://arxiv.org/abs/1011.6374} {arXiv:1011.6374 [astro-ph.CO]} \BibitemShut
  {NoStop}%
\bibitem [{\citenamefont {Zavala}\ \emph {et~al.}(2013)\citenamefont {Zavala},
  \citenamefont {Vogelsberger},\ and\ \citenamefont {Walker}}]{Zavala:2012us}%
  \BibitemOpen
  \bibfield  {author} {\bibinfo {author} {\bibfnamefont {J.}~\bibnamefont
  {Zavala}}, \bibinfo {author} {\bibfnamefont {M.}~\bibnamefont
  {Vogelsberger}}, \ and\ \bibinfo {author} {\bibfnamefont {M.~G.}\
  \bibnamefont {Walker}},\ }\href {\doibase 10.1093/mnrasl/sls053} {\bibfield
  {journal} {\bibinfo  {journal} {Monthly Notices of the Royal Astronomical
  Society: Letters}\ }\textbf {\bibinfo {volume} {431}},\ \bibinfo {pages}
  {L20} (\bibinfo {year} {2013})},\ \Eprint {http://arxiv.org/abs/1211.6426}
  {arXiv:1211.6426 [astro-ph.CO]} \BibitemShut {NoStop}%
\bibitem [{\citenamefont {Vogelsberger}\ \emph {et~al.}(2012)\citenamefont
  {Vogelsberger}, \citenamefont {Zavala},\ and\ \citenamefont
  {Loeb}}]{Vogelsberger:2012ku}%
  \BibitemOpen
  \bibfield  {author} {\bibinfo {author} {\bibfnamefont {M.}~\bibnamefont
  {Vogelsberger}}, \bibinfo {author} {\bibfnamefont {J.}~\bibnamefont
  {Zavala}}, \ and\ \bibinfo {author} {\bibfnamefont {A.}~\bibnamefont
  {Loeb}},\ }\href {\doibase 10.1111/j.1365-2966.2012.21182.x} {\bibfield
  {journal} {\bibinfo  {journal} {Mon. Not. Roy. Astron. Soc.}\ }\textbf
  {\bibinfo {volume} {423}},\ \bibinfo {pages} {3740} (\bibinfo {year}
  {2012})},\ \Eprint {http://arxiv.org/abs/1201.5892} {arXiv:1201.5892
  [astro-ph.CO]} \BibitemShut {NoStop}%
\bibitem [{\citenamefont {Markevitch}\ \emph {et~al.}(2004)\citenamefont
  {Markevitch}, \citenamefont {Gonzalez}, \citenamefont {Clowe}, \citenamefont
  {Vikhlinin}, \citenamefont {David}, \citenamefont {Forman}, \citenamefont
  {Jones}, \citenamefont {Murray},\ and\ \citenamefont
  {Tucker}}]{Markevitch:2003at}%
  \BibitemOpen
  \bibfield  {author} {\bibinfo {author} {\bibfnamefont {M.}~\bibnamefont
  {Markevitch}}, \bibinfo {author} {\bibfnamefont {A.~H.}\ \bibnamefont
  {Gonzalez}}, \bibinfo {author} {\bibfnamefont {D.}~\bibnamefont {Clowe}},
  \bibinfo {author} {\bibfnamefont {A.}~\bibnamefont {Vikhlinin}}, \bibinfo
  {author} {\bibfnamefont {L.}~\bibnamefont {David}}, \bibinfo {author}
  {\bibfnamefont {W.}~\bibnamefont {Forman}}, \bibinfo {author} {\bibfnamefont
  {C.}~\bibnamefont {Jones}}, \bibinfo {author} {\bibfnamefont
  {S.}~\bibnamefont {Murray}}, \ and\ \bibinfo {author} {\bibfnamefont
  {W.}~\bibnamefont {Tucker}},\ }\href {\doibase 10.1086/383178} {\bibfield
  {journal} {\bibinfo  {journal} {Astrophys. J.}\ }\textbf {\bibinfo {volume}
  {606}},\ \bibinfo {pages} {819} (\bibinfo {year} {2004})},\ \Eprint
  {http://arxiv.org/abs/astro-ph/0309303} {arXiv:astro-ph/0309303 [astro-ph]}
  \BibitemShut {NoStop}%
\bibitem [{\citenamefont {Randall}\ \emph {et~al.}(2008)\citenamefont
  {Randall}, \citenamefont {Markevitch}, \citenamefont {Clowe}, \citenamefont
  {Gonzalez},\ and\ \citenamefont {Bradac}}]{Randall:2007ph}%
  \BibitemOpen
  \bibfield  {author} {\bibinfo {author} {\bibfnamefont {S.~W.}\ \bibnamefont
  {Randall}}, \bibinfo {author} {\bibfnamefont {M.}~\bibnamefont {Markevitch}},
  \bibinfo {author} {\bibfnamefont {D.}~\bibnamefont {Clowe}}, \bibinfo
  {author} {\bibfnamefont {A.~H.}\ \bibnamefont {Gonzalez}}, \ and\ \bibinfo
  {author} {\bibfnamefont {M.}~\bibnamefont {Bradac}},\ }\href {\doibase
  10.1086/587859} {\bibfield  {journal} {\bibinfo  {journal} {Astrophys. J.}\
  }\textbf {\bibinfo {volume} {679}},\ \bibinfo {pages} {1173} (\bibinfo {year}
  {2008})},\ \Eprint {http://arxiv.org/abs/0704.0261} {arXiv:0704.0261
  [astro-ph]} \BibitemShut {NoStop}%
\bibitem [{\citenamefont {Peter}\ \emph {et~al.}(2013)\citenamefont {Peter},
  \citenamefont {Rocha}, \citenamefont {Bullock},\ and\ \citenamefont
  {Kaplinghat}}]{Peter:2012jh}%
  \BibitemOpen
  \bibfield  {author} {\bibinfo {author} {\bibfnamefont {A.~H.~G.}\
  \bibnamefont {Peter}}, \bibinfo {author} {\bibfnamefont {M.}~\bibnamefont
  {Rocha}}, \bibinfo {author} {\bibfnamefont {J.~S.}\ \bibnamefont {Bullock}},
  \ and\ \bibinfo {author} {\bibfnamefont {M.}~\bibnamefont {Kaplinghat}},\
  }\href {\doibase 10.1093/mnras/sts535} {\bibfield  {journal} {\bibinfo
  {journal} {Mon. Not. Roy. Astron. Soc.}\ }\textbf {\bibinfo {volume} {430}},\
  \bibinfo {pages} {105} (\bibinfo {year} {2013})},\ \Eprint
  {http://arxiv.org/abs/1208.3026} {arXiv:1208.3026 [astro-ph.CO]} \BibitemShut
  {NoStop}%
\bibitem [{\citenamefont {Rocha}\ \emph {et~al.}(2013)\citenamefont {Rocha},
  \citenamefont {Peter}, \citenamefont {Bullock}, \citenamefont {Kaplinghat},
  \citenamefont {Garrison-Kimmel}, \citenamefont {Onorbe},\ and\ \citenamefont
  {Moustakas}}]{Rocha:2012jg}%
  \BibitemOpen
  \bibfield  {author} {\bibinfo {author} {\bibfnamefont {M.}~\bibnamefont
  {Rocha}}, \bibinfo {author} {\bibfnamefont {A.~H.~G.}\ \bibnamefont {Peter}},
  \bibinfo {author} {\bibfnamefont {J.~S.}\ \bibnamefont {Bullock}}, \bibinfo
  {author} {\bibfnamefont {M.}~\bibnamefont {Kaplinghat}}, \bibinfo {author}
  {\bibfnamefont {S.}~\bibnamefont {Garrison-Kimmel}}, \bibinfo {author}
  {\bibfnamefont {J.}~\bibnamefont {Onorbe}}, \ and\ \bibinfo {author}
  {\bibfnamefont {L.~A.}\ \bibnamefont {Moustakas}},\ }\href {\doibase
  10.1093/mnras/sts514} {\bibfield  {journal} {\bibinfo  {journal} {Mon. Not.
  Roy. Astron. Soc.}\ }\textbf {\bibinfo {volume} {430}},\ \bibinfo {pages}
  {81} (\bibinfo {year} {2013})},\ \Eprint {http://arxiv.org/abs/1208.3025}
  {arXiv:1208.3025 [astro-ph.CO]} \BibitemShut {NoStop}%
\bibitem [{\citenamefont {Kahlhoefer}\ \emph {et~al.}(2014)\citenamefont
  {Kahlhoefer}, \citenamefont {Schmidt-Hoberg}, \citenamefont {Frandsen},\ and\
  \citenamefont {Sarkar}}]{Kahlhoefer:2013dca}%
  \BibitemOpen
  \bibfield  {author} {\bibinfo {author} {\bibfnamefont {F.}~\bibnamefont
  {Kahlhoefer}}, \bibinfo {author} {\bibfnamefont {K.}~\bibnamefont
  {Schmidt-Hoberg}}, \bibinfo {author} {\bibfnamefont {M.~T.}\ \bibnamefont
  {Frandsen}}, \ and\ \bibinfo {author} {\bibfnamefont {S.}~\bibnamefont
  {Sarkar}},\ }\href {\doibase 10.1093/mnras/stt2097} {\bibfield  {journal}
  {\bibinfo  {journal} {Mon. Not. Roy. Astron. Soc.}\ }\textbf {\bibinfo
  {volume} {437}},\ \bibinfo {pages} {2865} (\bibinfo {year} {2014})},\ \Eprint
  {http://arxiv.org/abs/1308.3419} {arXiv:1308.3419 [astro-ph.CO]} \BibitemShut
  {NoStop}%
\bibitem [{\citenamefont {Harvey}\ \emph {et~al.}(2015)\citenamefont {Harvey},
  \citenamefont {Massey}, \citenamefont {Kitching}, \citenamefont {Taylor},\
  and\ \citenamefont {Tittley}}]{Harvey:2015hha}%
  \BibitemOpen
  \bibfield  {author} {\bibinfo {author} {\bibfnamefont {D.}~\bibnamefont
  {Harvey}}, \bibinfo {author} {\bibfnamefont {R.}~\bibnamefont {Massey}},
  \bibinfo {author} {\bibfnamefont {T.}~\bibnamefont {Kitching}}, \bibinfo
  {author} {\bibfnamefont {A.}~\bibnamefont {Taylor}}, \ and\ \bibinfo {author}
  {\bibfnamefont {E.}~\bibnamefont {Tittley}},\ }\href {\doibase
  10.1126/science.1261381} {\bibfield  {journal} {\bibinfo  {journal}
  {Science}\ }\textbf {\bibinfo {volume} {347}},\ \bibinfo {pages} {1462}
  (\bibinfo {year} {2015})},\ \Eprint {http://arxiv.org/abs/1503.07675}
  {arXiv:1503.07675 [astro-ph.CO]} \BibitemShut {NoStop}%
\bibitem [{\citenamefont {Kaplinghat}\ \emph {et~al.}(2016)\citenamefont
  {Kaplinghat}, \citenamefont {Tulin},\ and\ \citenamefont
  {Yu}}]{Kaplinghat:2015aga}%
  \BibitemOpen
  \bibfield  {author} {\bibinfo {author} {\bibfnamefont {M.}~\bibnamefont
  {Kaplinghat}}, \bibinfo {author} {\bibfnamefont {S.}~\bibnamefont {Tulin}}, \
  and\ \bibinfo {author} {\bibfnamefont {H.-B.}\ \bibnamefont {Yu}},\ }\href
  {\doibase 10.1103/PhysRevLett.116.041302} {\bibfield  {journal} {\bibinfo
  {journal} {Phys. Rev. Lett.}\ }\textbf {\bibinfo {volume} {116}},\ \bibinfo
  {pages} {041302} (\bibinfo {year} {2016})},\ \Eprint
  {http://arxiv.org/abs/1508.03339} {arXiv:1508.03339 [astro-ph.CO]}
  \BibitemShut {NoStop}%
\bibitem [{\citenamefont {Ackerman}\ \emph {et~al.}(2009)\citenamefont
  {Ackerman}, \citenamefont {Buckley}, \citenamefont {Carroll},\ and\
  \citenamefont {Kamionkowski}}]{Ackerman:mha}%
  \BibitemOpen
  \bibfield  {author} {\bibinfo {author} {\bibfnamefont {L.}~\bibnamefont
  {Ackerman}}, \bibinfo {author} {\bibfnamefont {M.~R.}\ \bibnamefont
  {Buckley}}, \bibinfo {author} {\bibfnamefont {S.~M.}\ \bibnamefont
  {Carroll}}, \ and\ \bibinfo {author} {\bibfnamefont {M.}~\bibnamefont
  {Kamionkowski}},\ }\bibfield  {booktitle} {\emph {\bibinfo {booktitle}
  {{Proceedings, 7th International Heidelberg Conference on Dark Matter in
  Astro and Particle Physics (DARK 2009): Christchurch, New Zealand, January
  18-24, 2009}}},\ }\href {\doibase 10.1103/PhysRevD.79.023519,
  10.1142/9789814293792_0021} {\bibfield  {journal} {\bibinfo  {journal} {Phys.
  Rev.}\ }\textbf {\bibinfo {volume} {D79}},\ \bibinfo {pages} {023519}
  (\bibinfo {year} {2009})},\ \bibinfo {note} {[,277(2008)]},\ \Eprint
  {http://arxiv.org/abs/0810.5126} {arXiv:0810.5126 [hep-ph]} \BibitemShut
  {NoStop}%
\bibitem [{\citenamefont {van~den Aarssen}\ \emph
  {et~al.}(2012{\natexlab{a}})\citenamefont {van~den Aarssen}, \citenamefont
  {Bringmann},\ and\ \citenamefont {Pfrommer}}]{Aarssen:2012fx}%
  \BibitemOpen
  \bibfield  {author} {\bibinfo {author} {\bibfnamefont {L.~G.}\ \bibnamefont
  {van~den Aarssen}}, \bibinfo {author} {\bibfnamefont {T.}~\bibnamefont
  {Bringmann}}, \ and\ \bibinfo {author} {\bibfnamefont {C.}~\bibnamefont
  {Pfrommer}},\ }\href {\doibase 10.1103/PhysRevLett.109.231301} {\bibfield
  {journal} {\bibinfo  {journal} {Phys. Rev. Lett.}\ }\textbf {\bibinfo
  {volume} {109}},\ \bibinfo {pages} {231301} (\bibinfo {year}
  {2012}{\natexlab{a}})},\ \Eprint {http://arxiv.org/abs/1205.5809}
  {arXiv:1205.5809 [astro-ph.CO]} \BibitemShut {NoStop}%
\bibitem [{\citenamefont {Tulin}\ \emph {et~al.}(2013)\citenamefont {Tulin},
  \citenamefont {Yu},\ and\ \citenamefont {Zurek}}]{Tulin:2013teo}%
  \BibitemOpen
  \bibfield  {author} {\bibinfo {author} {\bibfnamefont {S.}~\bibnamefont
  {Tulin}}, \bibinfo {author} {\bibfnamefont {H.-B.}\ \bibnamefont {Yu}}, \
  and\ \bibinfo {author} {\bibfnamefont {K.~M.}\ \bibnamefont {Zurek}},\ }\href
  {\doibase 10.1103/PhysRevD.87.115007} {\bibfield  {journal} {\bibinfo
  {journal} {Phys. Rev.}\ }\textbf {\bibinfo {volume} {D87}},\ \bibinfo {pages}
  {115007} (\bibinfo {year} {2013})},\ \Eprint {http://arxiv.org/abs/1302.3898}
  {arXiv:1302.3898 [hep-ph]} \BibitemShut {NoStop}%
\bibitem [{\citenamefont {Gondolo}\ and\ \citenamefont
  {Gelmini}(1991)}]{Gondolo:1990dk}%
  \BibitemOpen
  \bibfield  {author} {\bibinfo {author} {\bibfnamefont {P.}~\bibnamefont
  {Gondolo}}\ and\ \bibinfo {author} {\bibfnamefont {G.}~\bibnamefont
  {Gelmini}},\ }\href {\doibase 10.1016/0550-3213(91)90438-4} {\bibfield
  {journal} {\bibinfo  {journal} {Nucl. Phys.}\ }\textbf {\bibinfo {volume}
  {B360}},\ \bibinfo {pages} {145} (\bibinfo {year} {1991})}\BibitemShut
  {NoStop}%
\bibitem [{\citenamefont {Chu}\ \emph {et~al.}(2016)\citenamefont {Chu},
  \citenamefont {Garcia-Cely},\ and\ \citenamefont {Hambye}}]{Chu:2016pew}%
  \BibitemOpen
  \bibfield  {author} {\bibinfo {author} {\bibfnamefont {X.}~\bibnamefont
  {Chu}}, \bibinfo {author} {\bibfnamefont {C.}~\bibnamefont {Garcia-Cely}}, \
  and\ \bibinfo {author} {\bibfnamefont {T.}~\bibnamefont {Hambye}},\ }\href
  {\doibase 10.1007/JHEP11(2016)048} {\bibfield  {journal} {\bibinfo  {journal}
  {JHEP}\ }\textbf {\bibinfo {volume} {11}},\ \bibinfo {pages} {048} (\bibinfo
  {year} {2016})},\ \Eprint {http://arxiv.org/abs/1609.00399} {arXiv:1609.00399
  [hep-ph]} \BibitemShut {NoStop}%
\bibitem [{\citenamefont {Pospelov}\ \emph {et~al.}(2008)\citenamefont
  {Pospelov}, \citenamefont {Ritz},\ and\ \citenamefont
  {Voloshin}}]{Pospelov:2007mp}%
  \BibitemOpen
  \bibfield  {author} {\bibinfo {author} {\bibfnamefont {M.}~\bibnamefont
  {Pospelov}}, \bibinfo {author} {\bibfnamefont {A.}~\bibnamefont {Ritz}}, \
  and\ \bibinfo {author} {\bibfnamefont {M.~B.}\ \bibnamefont {Voloshin}},\
  }\href {\doibase 10.1016/j.physletb.2008.02.052} {\bibfield  {journal}
  {\bibinfo  {journal} {Phys. Lett.}\ }\textbf {\bibinfo {volume} {B662}},\
  \bibinfo {pages} {53} (\bibinfo {year} {2008})},\ \Eprint
  {http://arxiv.org/abs/0711.4866} {arXiv:0711.4866 [hep-ph]} \BibitemShut
  {NoStop}%
\bibitem [{\citenamefont {Kaplinghat}\ \emph {et~al.}(2014)\citenamefont
  {Kaplinghat}, \citenamefont {Tulin},\ and\ \citenamefont
  {Yu}}]{Kaplinghat:2013yxa}%
  \BibitemOpen
  \bibfield  {author} {\bibinfo {author} {\bibfnamefont {M.}~\bibnamefont
  {Kaplinghat}}, \bibinfo {author} {\bibfnamefont {S.}~\bibnamefont {Tulin}}, \
  and\ \bibinfo {author} {\bibfnamefont {H.-B.}\ \bibnamefont {Yu}},\ }\href
  {\doibase 10.1103/PhysRevD.89.035009} {\bibfield  {journal} {\bibinfo
  {journal} {Phys. Rev.}\ }\textbf {\bibinfo {volume} {D89}},\ \bibinfo {pages}
  {035009} (\bibinfo {year} {2014})},\ \Eprint {http://arxiv.org/abs/1310.7945}
  {arXiv:1310.7945 [hep-ph]} \BibitemShut {NoStop}%
\bibitem [{\citenamefont {Del~Nobile}\ \emph {et~al.}(2015)\citenamefont
  {Del~Nobile}, \citenamefont {Kaplinghat},\ and\ \citenamefont
  {Yu}}]{DelNobile:2015uua}%
  \BibitemOpen
  \bibfield  {author} {\bibinfo {author} {\bibfnamefont {E.}~\bibnamefont
  {Del~Nobile}}, \bibinfo {author} {\bibfnamefont {M.}~\bibnamefont
  {Kaplinghat}}, \ and\ \bibinfo {author} {\bibfnamefont {H.-B.}\ \bibnamefont
  {Yu}},\ }\href {\doibase 10.1088/1475-7516/2015/10/055} {\bibfield  {journal}
  {\bibinfo  {journal} {JCAP}\ }\textbf {\bibinfo {volume} {1510}},\ \bibinfo
  {pages} {055} (\bibinfo {year} {2015})},\ \Eprint
  {http://arxiv.org/abs/1507.04007} {arXiv:1507.04007 [hep-ph]} \BibitemShut
  {NoStop}%
\bibitem [{\citenamefont {Kouvaris}\ \emph {et~al.}(2015)\citenamefont
  {Kouvaris}, \citenamefont {Shoemaker},\ and\ \citenamefont
  {Tuominen}}]{Kouvaris:2014uoa}%
  \BibitemOpen
  \bibfield  {author} {\bibinfo {author} {\bibfnamefont {C.}~\bibnamefont
  {Kouvaris}}, \bibinfo {author} {\bibfnamefont {I.~M.}\ \bibnamefont
  {Shoemaker}}, \ and\ \bibinfo {author} {\bibfnamefont {K.}~\bibnamefont
  {Tuominen}},\ }\href {\doibase 10.1103/PhysRevD.91.043519} {\bibfield
  {journal} {\bibinfo  {journal} {Phys. Rev.}\ }\textbf {\bibinfo {volume}
  {D91}},\ \bibinfo {pages} {043519} (\bibinfo {year} {2015})},\ \Eprint
  {http://arxiv.org/abs/1411.3730} {arXiv:1411.3730 [hep-ph]} \BibitemShut
  {NoStop}%
\bibitem [{\citenamefont {Bernal}\ \emph {et~al.}(2016)\citenamefont {Bernal},
  \citenamefont {Chu}, \citenamefont {Garcia-Cely}, \citenamefont {Hambye},\
  and\ \citenamefont {Zaldivar}}]{Bernal:2015ova}%
  \BibitemOpen
  \bibfield  {author} {\bibinfo {author} {\bibfnamefont {N.}~\bibnamefont
  {Bernal}}, \bibinfo {author} {\bibfnamefont {X.}~\bibnamefont {Chu}},
  \bibinfo {author} {\bibfnamefont {C.}~\bibnamefont {Garcia-Cely}}, \bibinfo
  {author} {\bibfnamefont {T.}~\bibnamefont {Hambye}}, \ and\ \bibinfo {author}
  {\bibfnamefont {B.}~\bibnamefont {Zaldivar}},\ }\href {\doibase
  10.1088/1475-7516/2016/03/018} {\bibfield  {journal} {\bibinfo  {journal}
  {JCAP}\ }\textbf {\bibinfo {volume} {1603}},\ \bibinfo {pages} {018}
  (\bibinfo {year} {2016})},\ \Eprint {http://arxiv.org/abs/1510.08063}
  {arXiv:1510.08063 [hep-ph]} \BibitemShut {NoStop}%
\bibitem [{\citenamefont {Kainulainen}\ \emph {et~al.}(2016)\citenamefont
  {Kainulainen}, \citenamefont {Tuominen},\ and\ \citenamefont
  {Vaskonen}}]{Kainulainen:2015sva}%
  \BibitemOpen
  \bibfield  {author} {\bibinfo {author} {\bibfnamefont {K.}~\bibnamefont
  {Kainulainen}}, \bibinfo {author} {\bibfnamefont {K.}~\bibnamefont
  {Tuominen}}, \ and\ \bibinfo {author} {\bibfnamefont {V.}~\bibnamefont
  {Vaskonen}},\ }\href {\doibase 10.1103/PhysRevD.93.015016} {\bibfield
  {journal} {\bibinfo  {journal} {Phys. Rev.}\ }\textbf {\bibinfo {volume}
  {D93}},\ \bibinfo {pages} {015016} (\bibinfo {year} {2016})},\ \Eprint
  {http://arxiv.org/abs/1507.04931} {arXiv:1507.04931 [hep-ph]} \BibitemShut
  {NoStop}%
\bibitem [{\citenamefont {Kamionkowski}\ and\ \citenamefont
  {Profumo}(2008)}]{Kamionkowski:2008gj}%
  \BibitemOpen
  \bibfield  {author} {\bibinfo {author} {\bibfnamefont {M.}~\bibnamefont
  {Kamionkowski}}\ and\ \bibinfo {author} {\bibfnamefont {S.}~\bibnamefont
  {Profumo}},\ }\href {\doibase 10.1103/PhysRevLett.101.261301} {\bibfield
  {journal} {\bibinfo  {journal} {Phys. Rev. Lett.}\ }\textbf {\bibinfo
  {volume} {101}},\ \bibinfo {pages} {261301} (\bibinfo {year} {2008})},\
  \Eprint {http://arxiv.org/abs/0810.3233} {arXiv:0810.3233 [astro-ph]}
  \BibitemShut {NoStop}%
\bibitem [{\citenamefont {Zavala}\ \emph {et~al.}(2010)\citenamefont {Zavala},
  \citenamefont {Vogelsberger},\ and\ \citenamefont {White}}]{Zavala:2009mi}%
  \BibitemOpen
  \bibfield  {author} {\bibinfo {author} {\bibfnamefont {J.}~\bibnamefont
  {Zavala}}, \bibinfo {author} {\bibfnamefont {M.}~\bibnamefont
  {Vogelsberger}}, \ and\ \bibinfo {author} {\bibfnamefont {S.~D.~M.}\
  \bibnamefont {White}},\ }\href {\doibase 10.1103/PhysRevD.81.083502}
  {\bibfield  {journal} {\bibinfo  {journal} {Phys. Rev.}\ }\textbf {\bibinfo
  {volume} {D81}},\ \bibinfo {pages} {083502} (\bibinfo {year} {2010})},\
  \Eprint {http://arxiv.org/abs/0910.5221} {arXiv:0910.5221 [astro-ph.CO]}
  \BibitemShut {NoStop}%
\bibitem [{\citenamefont {Feng}\ \emph
  {et~al.}(2010{\natexlab{b}})\citenamefont {Feng}, \citenamefont
  {Kaplinghat},\ and\ \citenamefont {Yu}}]{Feng:2010zp}%
  \BibitemOpen
  \bibfield  {author} {\bibinfo {author} {\bibfnamefont {J.~L.}\ \bibnamefont
  {Feng}}, \bibinfo {author} {\bibfnamefont {M.}~\bibnamefont {Kaplinghat}}, \
  and\ \bibinfo {author} {\bibfnamefont {H.-B.}\ \bibnamefont {Yu}},\ }\href
  {\doibase 10.1103/PhysRevD.82.083525} {\bibfield  {journal} {\bibinfo
  {journal} {Phys. Rev.}\ }\textbf {\bibinfo {volume} {D82}},\ \bibinfo {pages}
  {083525} (\bibinfo {year} {2010}{\natexlab{b}})},\ \Eprint
  {http://arxiv.org/abs/1005.4678} {arXiv:1005.4678 [hep-ph]} \BibitemShut
  {NoStop}%
\bibitem [{\citenamefont {Hisano}\ \emph {et~al.}(2011)\citenamefont {Hisano},
  \citenamefont {Kawasaki}, \citenamefont {Kohri}, \citenamefont {Moroi},
  \citenamefont {Nakayama},\ and\ \citenamefont {Sekiguchi}}]{Hisano:2011dc}%
  \BibitemOpen
  \bibfield  {author} {\bibinfo {author} {\bibfnamefont {J.}~\bibnamefont
  {Hisano}}, \bibinfo {author} {\bibfnamefont {M.}~\bibnamefont {Kawasaki}},
  \bibinfo {author} {\bibfnamefont {K.}~\bibnamefont {Kohri}}, \bibinfo
  {author} {\bibfnamefont {T.}~\bibnamefont {Moroi}}, \bibinfo {author}
  {\bibfnamefont {K.}~\bibnamefont {Nakayama}}, \ and\ \bibinfo {author}
  {\bibfnamefont {T.}~\bibnamefont {Sekiguchi}},\ }\href {\doibase
  10.1103/PhysRevD.83.123511} {\bibfield  {journal} {\bibinfo  {journal} {Phys.
  Rev.}\ }\textbf {\bibinfo {volume} {D83}},\ \bibinfo {pages} {123511}
  (\bibinfo {year} {2011})},\ \Eprint {http://arxiv.org/abs/1102.4658}
  {arXiv:1102.4658 [hep-ph]} \BibitemShut {NoStop}%
\bibitem [{\citenamefont {Arkani-Hamed}\ \emph {et~al.}(2009)\citenamefont
  {Arkani-Hamed}, \citenamefont {Finkbeiner}, \citenamefont {Slatyer},\ and\
  \citenamefont {Weiner}}]{ArkaniHamed:2008qn}%
  \BibitemOpen
  \bibfield  {author} {\bibinfo {author} {\bibfnamefont {N.}~\bibnamefont
  {Arkani-Hamed}}, \bibinfo {author} {\bibfnamefont {D.~P.}\ \bibnamefont
  {Finkbeiner}}, \bibinfo {author} {\bibfnamefont {T.~R.}\ \bibnamefont
  {Slatyer}}, \ and\ \bibinfo {author} {\bibfnamefont {N.}~\bibnamefont
  {Weiner}},\ }\href {\doibase 10.1103/PhysRevD.79.015014} {\bibfield
  {journal} {\bibinfo  {journal} {Phys. Rev.}\ }\textbf {\bibinfo {volume}
  {D79}},\ \bibinfo {pages} {015014} (\bibinfo {year} {2009})},\ \Eprint
  {http://arxiv.org/abs/0810.0713} {arXiv:0810.0713 [hep-ph]} \BibitemShut
  {NoStop}%
\bibitem [{\citenamefont {Bergstrom}\ \emph {et~al.}(2009)\citenamefont
  {Bergstrom}, \citenamefont {Bertone}, \citenamefont {Bringmann},
  \citenamefont {Edsjo},\ and\ \citenamefont {Taoso}}]{Bergstrom:2008ag}%
  \BibitemOpen
  \bibfield  {author} {\bibinfo {author} {\bibfnamefont {L.}~\bibnamefont
  {Bergstrom}}, \bibinfo {author} {\bibfnamefont {G.}~\bibnamefont {Bertone}},
  \bibinfo {author} {\bibfnamefont {T.}~\bibnamefont {Bringmann}}, \bibinfo
  {author} {\bibfnamefont {J.}~\bibnamefont {Edsjo}}, \ and\ \bibinfo {author}
  {\bibfnamefont {M.}~\bibnamefont {Taoso}},\ }\href {\doibase
  10.1103/PhysRevD.79.081303} {\bibfield  {journal} {\bibinfo  {journal} {Phys.
  Rev.}\ }\textbf {\bibinfo {volume} {D79}},\ \bibinfo {pages} {081303}
  (\bibinfo {year} {2009})},\ \Eprint {http://arxiv.org/abs/0812.3895}
  {arXiv:0812.3895 [astro-ph]} \BibitemShut {NoStop}%
\bibitem [{\citenamefont {Mardon}\ \emph {et~al.}(2009)\citenamefont {Mardon},
  \citenamefont {Nomura}, \citenamefont {Stolarski},\ and\ \citenamefont
  {Thaler}}]{Mardon:2009rc}%
  \BibitemOpen
  \bibfield  {author} {\bibinfo {author} {\bibfnamefont {J.}~\bibnamefont
  {Mardon}}, \bibinfo {author} {\bibfnamefont {Y.}~\bibnamefont {Nomura}},
  \bibinfo {author} {\bibfnamefont {D.}~\bibnamefont {Stolarski}}, \ and\
  \bibinfo {author} {\bibfnamefont {J.}~\bibnamefont {Thaler}},\ }\href
  {\doibase 10.1088/1475-7516/2009/05/016} {\bibfield  {journal} {\bibinfo
  {journal} {JCAP}\ }\textbf {\bibinfo {volume} {0905}},\ \bibinfo {pages}
  {016} (\bibinfo {year} {2009})},\ \Eprint {http://arxiv.org/abs/0901.2926}
  {arXiv:0901.2926 [hep-ph]} \BibitemShut {NoStop}%
\bibitem [{\citenamefont {Galli}\ \emph {et~al.}(2009)\citenamefont {Galli},
  \citenamefont {Iocco}, \citenamefont {Bertone},\ and\ \citenamefont
  {Melchiorri}}]{Galli:2009zc}%
  \BibitemOpen
  \bibfield  {author} {\bibinfo {author} {\bibfnamefont {S.}~\bibnamefont
  {Galli}}, \bibinfo {author} {\bibfnamefont {F.}~\bibnamefont {Iocco}},
  \bibinfo {author} {\bibfnamefont {G.}~\bibnamefont {Bertone}}, \ and\
  \bibinfo {author} {\bibfnamefont {A.}~\bibnamefont {Melchiorri}},\ }\href
  {\doibase 10.1103/PhysRevD.80.023505} {\bibfield  {journal} {\bibinfo
  {journal} {Phys. Rev.}\ }\textbf {\bibinfo {volume} {D80}},\ \bibinfo {pages}
  {023505} (\bibinfo {year} {2009})},\ \Eprint {http://arxiv.org/abs/0905.0003}
  {arXiv:0905.0003 [astro-ph.CO]} \BibitemShut {NoStop}%
\bibitem [{\citenamefont {Slatyer}\ \emph {et~al.}(2009)\citenamefont
  {Slatyer}, \citenamefont {Padmanabhan},\ and\ \citenamefont
  {Finkbeiner}}]{Slatyer:2009yq}%
  \BibitemOpen
  \bibfield  {author} {\bibinfo {author} {\bibfnamefont {T.~R.}\ \bibnamefont
  {Slatyer}}, \bibinfo {author} {\bibfnamefont {N.}~\bibnamefont
  {Padmanabhan}}, \ and\ \bibinfo {author} {\bibfnamefont {D.~P.}\ \bibnamefont
  {Finkbeiner}},\ }\href {\doibase 10.1103/PhysRevD.80.043526} {\bibfield
  {journal} {\bibinfo  {journal} {Phys. Rev.}\ }\textbf {\bibinfo {volume}
  {D80}},\ \bibinfo {pages} {043526} (\bibinfo {year} {2009})},\ \Eprint
  {http://arxiv.org/abs/0906.1197} {arXiv:0906.1197 [astro-ph.CO]} \BibitemShut
  {NoStop}%
\bibitem [{\citenamefont {Hannestad}\ and\ \citenamefont
  {Tram}(2011)}]{Hannestad:2010zt}%
  \BibitemOpen
  \bibfield  {author} {\bibinfo {author} {\bibfnamefont {S.}~\bibnamefont
  {Hannestad}}\ and\ \bibinfo {author} {\bibfnamefont {T.}~\bibnamefont
  {Tram}},\ }\href {\doibase 10.1088/1475-7516/2011/01/016} {\bibfield
  {journal} {\bibinfo  {journal} {JCAP}\ }\textbf {\bibinfo {volume} {1101}},\
  \bibinfo {pages} {016} (\bibinfo {year} {2011})},\ \Eprint
  {http://arxiv.org/abs/1008.1511} {arXiv:1008.1511 [astro-ph.CO]} \BibitemShut
  {NoStop}%
\bibitem [{\citenamefont {Finkbeiner}\ \emph {et~al.}(2011)\citenamefont
  {Finkbeiner}, \citenamefont {Goodenough}, \citenamefont {Slatyer},
  \citenamefont {Vogelsberger},\ and\ \citenamefont
  {Weiner}}]{Finkbeiner:2010sm}%
  \BibitemOpen
  \bibfield  {author} {\bibinfo {author} {\bibfnamefont {D.~P.}\ \bibnamefont
  {Finkbeiner}}, \bibinfo {author} {\bibfnamefont {L.}~\bibnamefont
  {Goodenough}}, \bibinfo {author} {\bibfnamefont {T.~R.}\ \bibnamefont
  {Slatyer}}, \bibinfo {author} {\bibfnamefont {M.}~\bibnamefont
  {Vogelsberger}}, \ and\ \bibinfo {author} {\bibfnamefont {N.}~\bibnamefont
  {Weiner}},\ }\href {\doibase 10.1088/1475-7516/2011/05/002} {\bibfield
  {journal} {\bibinfo  {journal} {JCAP}\ }\textbf {\bibinfo {volume} {1105}},\
  \bibinfo {pages} {002} (\bibinfo {year} {2011})},\ \Eprint
  {http://arxiv.org/abs/1011.3082} {arXiv:1011.3082 [hep-ph]} \BibitemShut
  {NoStop}%
\bibitem [{\citenamefont {Cyr-Racine}\ \emph {et~al.}(2016)\citenamefont
  {Cyr-Racine}, \citenamefont {Sigurdson}, \citenamefont {Zavala},
  \citenamefont {Bringmann}, \citenamefont {Vogelsberger},\ and\ \citenamefont
  {Pfrommer}}]{Cyr-Racine:2015ihg}%
  \BibitemOpen
  \bibfield  {author} {\bibinfo {author} {\bibfnamefont {F.-Y.}\ \bibnamefont
  {Cyr-Racine}}, \bibinfo {author} {\bibfnamefont {K.}~\bibnamefont
  {Sigurdson}}, \bibinfo {author} {\bibfnamefont {J.}~\bibnamefont {Zavala}},
  \bibinfo {author} {\bibfnamefont {T.}~\bibnamefont {Bringmann}}, \bibinfo
  {author} {\bibfnamefont {M.}~\bibnamefont {Vogelsberger}}, \ and\ \bibinfo
  {author} {\bibfnamefont {C.}~\bibnamefont {Pfrommer}},\ }\href {\doibase
  10.1103/PhysRevD.93.123527} {\bibfield  {journal} {\bibinfo  {journal} {Phys.
  Rev.}\ }\textbf {\bibinfo {volume} {D93}},\ \bibinfo {pages} {123527}
  (\bibinfo {year} {2016})},\ \Eprint {http://arxiv.org/abs/1512.05344}
  {arXiv:1512.05344 [astro-ph.CO]} \BibitemShut {NoStop}%
\bibitem [{\citenamefont {de~Blok}\ and\ \citenamefont
  {McGaugh}(1997)}]{deBlok:1997zlw}%
  \BibitemOpen
  \bibfield  {author} {\bibinfo {author} {\bibfnamefont {W.~J.~G.}\
  \bibnamefont {de~Blok}}\ and\ \bibinfo {author} {\bibfnamefont {S.~S.}\
  \bibnamefont {McGaugh}},\ }\href {\doibase 10.1093/mnras/290.3.533}
  {\bibfield  {journal} {\bibinfo  {journal} {Mon. Not. Roy. Astron. Soc.}\
  }\textbf {\bibinfo {volume} {290}},\ \bibinfo {pages} {533} (\bibinfo {year}
  {1997})},\ \Eprint {http://arxiv.org/abs/astro-ph/9704274}
  {arXiv:astro-ph/9704274 [astro-ph]} \BibitemShut {NoStop}%
\bibitem [{\citenamefont {Oh}\ \emph {et~al.}(2011)\citenamefont {Oh},
  \citenamefont {de~Blok}, \citenamefont {Brinks}, \citenamefont {Walter},\
  and\ \citenamefont {Kennicutt}}]{Oh:2010ea}%
  \BibitemOpen
  \bibfield  {author} {\bibinfo {author} {\bibfnamefont {S.-H.}\ \bibnamefont
  {Oh}}, \bibinfo {author} {\bibfnamefont {W.~J.~G.}\ \bibnamefont {de~Blok}},
  \bibinfo {author} {\bibfnamefont {E.}~\bibnamefont {Brinks}}, \bibinfo
  {author} {\bibfnamefont {F.}~\bibnamefont {Walter}}, \ and\ \bibinfo {author}
  {\bibfnamefont {R.~C.}\ \bibnamefont {Kennicutt}, \bibfnamefont {Jr}},\
  }\href {\doibase 10.1088/0004-6256/141/6/193} {\bibfield  {journal} {\bibinfo
   {journal} {Astron. J.}\ }\textbf {\bibinfo {volume} {141}},\ \bibinfo
  {pages} {193} (\bibinfo {year} {2011})},\ \Eprint
  {http://arxiv.org/abs/1011.0899} {arXiv:1011.0899 [astro-ph.CO]} \BibitemShut
  {NoStop}%
\bibitem [{\citenamefont {Walker}\ and\ \citenamefont
  {Penarrubia}(2011)}]{Walker:2011zu}%
  \BibitemOpen
  \bibfield  {author} {\bibinfo {author} {\bibfnamefont {M.~G.}\ \bibnamefont
  {Walker}}\ and\ \bibinfo {author} {\bibfnamefont {J.}~\bibnamefont
  {Penarrubia}},\ }\href {\doibase 10.1088/0004-637X/742/1/20} {\bibfield
  {journal} {\bibinfo  {journal} {Astrophys. J.}\ }\textbf {\bibinfo {volume}
  {742}},\ \bibinfo {pages} {20} (\bibinfo {year} {2011})},\ \Eprint
  {http://arxiv.org/abs/1108.2404} {arXiv:1108.2404 [astro-ph.CO]} \BibitemShut
  {NoStop}%
\bibitem [{\citenamefont {Boylan-Kolchin}\ \emph {et~al.}(2011)\citenamefont
  {Boylan-Kolchin}, \citenamefont {Bullock},\ and\ \citenamefont
  {Kaplinghat}}]{BoylanKolchin:2011de}%
  \BibitemOpen
  \bibfield  {author} {\bibinfo {author} {\bibfnamefont {M.}~\bibnamefont
  {Boylan-Kolchin}}, \bibinfo {author} {\bibfnamefont {J.~S.}\ \bibnamefont
  {Bullock}}, \ and\ \bibinfo {author} {\bibfnamefont {M.}~\bibnamefont
  {Kaplinghat}},\ }\href {\doibase 10.1111/j.1745-3933.2011.01074.x} {\bibfield
   {journal} {\bibinfo  {journal} {Mon. Not. Roy. Astron. Soc.}\ }\textbf
  {\bibinfo {volume} {415}},\ \bibinfo {pages} {L40} (\bibinfo {year}
  {2011})},\ \Eprint {http://arxiv.org/abs/1103.0007} {arXiv:1103.0007
  [astro-ph.CO]} \BibitemShut {NoStop}%
\bibitem [{\citenamefont {Papastergis}\ \emph {et~al.}(2015)\citenamefont
  {Papastergis}, \citenamefont {Giovanelli}, \citenamefont {Haynes},\ and\
  \citenamefont {Shankar}}]{Papastergis:2014aba}%
  \BibitemOpen
  \bibfield  {author} {\bibinfo {author} {\bibfnamefont {E.}~\bibnamefont
  {Papastergis}}, \bibinfo {author} {\bibfnamefont {R.}~\bibnamefont
  {Giovanelli}}, \bibinfo {author} {\bibfnamefont {M.~P.}\ \bibnamefont
  {Haynes}}, \ and\ \bibinfo {author} {\bibfnamefont {F.}~\bibnamefont
  {Shankar}},\ }\href {\doibase 10.1051/0004-6361/201424909} {\bibfield
  {journal} {\bibinfo  {journal} {Astron. Astrophys.}\ }\textbf {\bibinfo
  {volume} {574}},\ \bibinfo {pages} {A113} (\bibinfo {year} {2015})},\ \Eprint
  {http://arxiv.org/abs/1407.4665} {arXiv:1407.4665 [astro-ph.GA]} \BibitemShut
  {NoStop}%
\bibitem [{\citenamefont {Ade}\ \emph {et~al.}(2016)\citenamefont {Ade} \emph
  {et~al.}}]{Ade:2015xua}%
  \BibitemOpen
  \bibfield  {author} {\bibinfo {author} {\bibfnamefont {P.~A.~R.}\
  \bibnamefont {Ade}} \emph {et~al.} (\bibinfo {collaboration} {Planck}),\
  }\href {\doibase 10.1051/0004-6361/201525830} {\bibfield  {journal} {\bibinfo
   {journal} {Astron. Astrophys.}\ }\textbf {\bibinfo {volume} {594}},\
  \bibinfo {pages} {A13} (\bibinfo {year} {2016})},\ \Eprint
  {http://arxiv.org/abs/1502.01589} {arXiv:1502.01589 [astro-ph.CO]}
  \BibitemShut {NoStop}%
\bibitem [{\citenamefont {Sommerfeld}(1931)}]{Sommerfeld}%
  \BibitemOpen
  \bibfield  {author} {\bibinfo {author} {\bibfnamefont {A.}~\bibnamefont
  {Sommerfeld}},\ }\href@noop {} {\bibfield  {journal} {\bibinfo  {journal}
  {Annalen der Physik}\ }\textbf {\bibinfo {volume} {403}},\ \bibinfo {pages}
  {207} (\bibinfo {year} {1931})}\BibitemShut {NoStop}%
\bibitem [{\citenamefont {Cassel}(2010)}]{Cassel:2009wt}%
  \BibitemOpen
  \bibfield  {author} {\bibinfo {author} {\bibfnamefont {S.}~\bibnamefont
  {Cassel}},\ }\href {\doibase 10.1088/0954-3899/37/10/105009} {\bibfield
  {journal} {\bibinfo  {journal} {J. Phys.}\ }\textbf {\bibinfo {volume}
  {G37}},\ \bibinfo {pages} {105009} (\bibinfo {year} {2010})},\ \Eprint
  {http://arxiv.org/abs/0903.5307} {arXiv:0903.5307 [hep-ph]} \BibitemShut
  {NoStop}%
\bibitem [{\citenamefont {Iengo}(2009)}]{Iengo:2009ni}%
  \BibitemOpen
  \bibfield  {author} {\bibinfo {author} {\bibfnamefont {R.}~\bibnamefont
  {Iengo}},\ }\href {\doibase 10.1088/1126-6708/2009/05/024} {\bibfield
  {journal} {\bibinfo  {journal} {JHEP}\ }\textbf {\bibinfo {volume} {05}},\
  \bibinfo {pages} {024} (\bibinfo {year} {2009})},\ \Eprint
  {http://arxiv.org/abs/0902.0688} {arXiv:0902.0688 [hep-ph]} \BibitemShut
  {NoStop}%
\bibitem [{\citenamefont {Slatyer}(2010)}]{Slatyer:2009vg}%
  \BibitemOpen
  \bibfield  {author} {\bibinfo {author} {\bibfnamefont {T.~R.}\ \bibnamefont
  {Slatyer}},\ }\href {\doibase 10.1088/1475-7516/2010/02/028} {\bibfield
  {journal} {\bibinfo  {journal} {JCAP}\ }\textbf {\bibinfo {volume} {1002}},\
  \bibinfo {pages} {028} (\bibinfo {year} {2010})},\ \Eprint
  {http://arxiv.org/abs/0910.5713} {arXiv:0910.5713 [hep-ph]} \BibitemShut
  {NoStop}%
\bibitem [{\citenamefont {Dent}\ \emph {et~al.}(2010)\citenamefont {Dent},
  \citenamefont {Dutta},\ and\ \citenamefont {Scherrer}}]{Dent:2009bv}%
  \BibitemOpen
  \bibfield  {author} {\bibinfo {author} {\bibfnamefont {J.~B.}\ \bibnamefont
  {Dent}}, \bibinfo {author} {\bibfnamefont {S.}~\bibnamefont {Dutta}}, \ and\
  \bibinfo {author} {\bibfnamefont {R.~J.}\ \bibnamefont {Scherrer}},\ }\href
  {\doibase 10.1016/j.physletb.2010.03.018} {\bibfield  {journal} {\bibinfo
  {journal} {Phys. Lett.}\ }\textbf {\bibinfo {volume} {B687}},\ \bibinfo
  {pages} {275} (\bibinfo {year} {2010})},\ \Eprint
  {http://arxiv.org/abs/0909.4128} {arXiv:0909.4128 [astro-ph.CO]} \BibitemShut
  {NoStop}%
\bibitem [{\citenamefont {van~den Aarssen}\ \emph
  {et~al.}(2012{\natexlab{b}})\citenamefont {van~den Aarssen}, \citenamefont
  {Bringmann},\ and\ \citenamefont {Goedecke}}]{vandenAarssen:2012ag}%
  \BibitemOpen
  \bibfield  {author} {\bibinfo {author} {\bibfnamefont {L.~G.}\ \bibnamefont
  {van~den Aarssen}}, \bibinfo {author} {\bibfnamefont {T.}~\bibnamefont
  {Bringmann}}, \ and\ \bibinfo {author} {\bibfnamefont {Y.~C.}\ \bibnamefont
  {Goedecke}},\ }\href {\doibase 10.1103/PhysRevD.85.123512} {\bibfield
  {journal} {\bibinfo  {journal} {Phys. Rev.}\ }\textbf {\bibinfo {volume}
  {D85}},\ \bibinfo {pages} {123512} (\bibinfo {year} {2012}{\natexlab{b}})},\
  \Eprint {http://arxiv.org/abs/1202.5456} {arXiv:1202.5456 [hep-ph]}
  \BibitemShut {NoStop}%
\bibitem [{\citenamefont {von Harling}\ and\ \citenamefont
  {Petraki}(2014)}]{vonHarling:2014kha}%
  \BibitemOpen
  \bibfield  {author} {\bibinfo {author} {\bibfnamefont {B.}~\bibnamefont {von
  Harling}}\ and\ \bibinfo {author} {\bibfnamefont {K.}~\bibnamefont
  {Petraki}},\ }\href {\doibase 10.1088/1475-7516/2014/12/033} {\bibfield
  {journal} {\bibinfo  {journal} {JCAP}\ }\textbf {\bibinfo {volume} {1412}},\
  \bibinfo {pages} {033} (\bibinfo {year} {2014})},\ \Eprint
  {http://arxiv.org/abs/1407.7874} {arXiv:1407.7874 [hep-ph]} \BibitemShut
  {NoStop}%
\bibitem [{\citenamefont {Cirelli}\ \emph {et~al.}(2016)\citenamefont
  {Cirelli}, \citenamefont {Panci}, \citenamefont {Petraki},\ 
  \citenamefont {Sala},\ and\ \citenamefont {Taoso}}]{Cirelli:2016rnw}%
  \BibitemOpen
  \bibfield  {author} {\bibinfo {author} {\bibfnamefont {M.}~\bibnamefont
  {Cirelli}}, \bibinfo {author} {\bibfnamefont {P.}~\bibnamefont
  {Panci}}, \bibinfo {author} {\bibfnamefont {K.}\ \bibnamefont
  {Petraki}}, \bibinfo {author} {\bibfnamefont {F.}\ \bibnamefont
  {Sala}}, \ and\ \bibinfo {author} {\bibfnamefont {M.}~\bibnamefont
  {Taoso}},\ }  {\bibfield {journal} \bibinfo {year} {2016}},\
  \Eprint {http://arxiv.org/abs/1612.07295} {arXiv:1612.07295 [hep-ph]} \BibitemShut
  {NoStop}%
\bibitem [{\citenamefont {Adams}\ \emph {et~al.}(1998)\citenamefont {Adams},
  \citenamefont {Sarkar},\ and\ \citenamefont {Sciama}}]{Adams:1998nr}%
  \BibitemOpen
  \bibfield  {author} {\bibinfo {author} {\bibfnamefont {J.~A.}\ \bibnamefont
  {Adams}}, \bibinfo {author} {\bibfnamefont {S.}~\bibnamefont {Sarkar}}, \
  and\ \bibinfo {author} {\bibfnamefont {D.~W.}\ \bibnamefont {Sciama}},\
  }\href {\doibase 10.1046/j.1365-8711.1998.02017.x} {\bibfield  {journal}
  {\bibinfo  {journal} {Mon. Not. Roy. Astron. Soc.}\ }\textbf {\bibinfo
  {volume} {301}},\ \bibinfo {pages} {210} (\bibinfo {year} {1998})},\ \Eprint
  {http://arxiv.org/abs/astro-ph/9805108} {arXiv:astro-ph/9805108 [astro-ph]}
  \BibitemShut {NoStop}%
\bibitem [{\citenamefont {Chen}\ and\ \citenamefont
  {Kamionkowski}(2004)}]{Chen:2003gz}%
  \BibitemOpen
  \bibfield  {author} {\bibinfo {author} {\bibfnamefont {X.-L.}\ \bibnamefont
  {Chen}}\ and\ \bibinfo {author} {\bibfnamefont {M.}~\bibnamefont
  {Kamionkowski}},\ }\href {\doibase 10.1103/PhysRevD.70.043502} {\bibfield
  {journal} {\bibinfo  {journal} {Phys. Rev.}\ }\textbf {\bibinfo {volume}
  {D70}},\ \bibinfo {pages} {043502} (\bibinfo {year} {2004})},\ \Eprint
  {http://arxiv.org/abs/astro-ph/0310473} {arXiv:astro-ph/0310473 [astro-ph]}
  \BibitemShut {NoStop}%
\bibitem [{\citenamefont {Cline}\ and\ \citenamefont
  {Scott}(2013)}]{Cline:2013fm}%
  \BibitemOpen
  \bibfield  {author} {\bibinfo {author} {\bibfnamefont {J.~M.}\ \bibnamefont
  {Cline}}\ and\ \bibinfo {author} {\bibfnamefont {P.}~\bibnamefont {Scott}},\
  }\href {\doibase 10.1088/1475-7516/2013/03/044,
  10.1088/1475-7516/2013/05/E01} {\bibfield  {journal} {\bibinfo  {journal}
  {JCAP}\ }\textbf {\bibinfo {volume} {1303}},\ \bibinfo {pages} {044}
  (\bibinfo {year} {2013})},\ \bibinfo {note} {[Erratum: JCAP1305,E01(2013)]},\
  \Eprint {http://arxiv.org/abs/1301.5908} {arXiv:1301.5908 [astro-ph.CO]}
  \BibitemShut {NoStop}%
\bibitem [{\citenamefont {Liu}\ \emph {et~al.}(2016)\citenamefont {Liu},
  \citenamefont {Slatyer},\ and\ \citenamefont {Zavala}}]{Liu:2016cnk}%
  \BibitemOpen
  \bibfield  {author} {\bibinfo {author} {\bibfnamefont {H.}~\bibnamefont
  {Liu}}, \bibinfo {author} {\bibfnamefont {T.~R.}\ \bibnamefont {Slatyer}}, \
  and\ \bibinfo {author} {\bibfnamefont {J.}~\bibnamefont {Zavala}},\ }\href
  {\doibase 10.1103/PhysRevD.94.063507} {\bibfield  {journal} {\bibinfo
  {journal} {Phys. Rev.}\ }\textbf {\bibinfo {volume} {D94}},\ \bibinfo {pages}
  {063507} (\bibinfo {year} {2016})},\ \Eprint
  {http://arxiv.org/abs/1604.02457} {arXiv:1604.02457 [astro-ph.CO]}
  \BibitemShut {NoStop}%
\bibitem [{\citenamefont {Slatyer}(2016)}]{Slatyer:2015jla}%
  \BibitemOpen
  \bibfield  {author} {\bibinfo {author} {\bibfnamefont {T.~R.}\ \bibnamefont
  {Slatyer}},\ }\href {\doibase 10.1103/PhysRevD.93.023527} {\bibfield
  {journal} {\bibinfo  {journal} {Phys. Rev.}\ }\textbf {\bibinfo {volume}
  {D93}},\ \bibinfo {pages} {023527} (\bibinfo {year} {2016})},\ \Eprint
  {http://arxiv.org/abs/1506.03811} {arXiv:1506.03811 [hep-ph]} \BibitemShut
  {NoStop}%
\bibitem [{\citenamefont {Bringmann}(2009)}]{Bringmann:2009vf}%
  \BibitemOpen
  \bibfield  {author} {\bibinfo {author} {\bibfnamefont {T.}~\bibnamefont
  {Bringmann}},\ }\href {\doibase 10.1088/1367-2630/11/10/105027} {\bibfield
  {journal} {\bibinfo  {journal} {New J. Phys.}\ }\textbf {\bibinfo {volume}
  {11}},\ \bibinfo {pages} {105027} (\bibinfo {year} {2009})},\ \Eprint
  {http://arxiv.org/abs/0903.0189} {arXiv:0903.0189 [astro-ph.CO]} \BibitemShut
  {NoStop}%
\bibitem [{\citenamefont {Croft}\ \emph {et~al.}(1998)\citenamefont {Croft},
  \citenamefont {Weinberg}, \citenamefont {Katz},\ and\ \citenamefont
  {Hernquist}}]{Croft:1997jf}%
  \BibitemOpen
  \bibfield  {author} {\bibinfo {author} {\bibfnamefont {R.~A.~C.}\
  \bibnamefont {Croft}}, \bibinfo {author} {\bibfnamefont {D.~H.}\ \bibnamefont
  {Weinberg}}, \bibinfo {author} {\bibfnamefont {N.}~\bibnamefont {Katz}}, \
  and\ \bibinfo {author} {\bibfnamefont {L.}~\bibnamefont {Hernquist}},\ }\href
  {\doibase 10.1086/305289} {\bibfield  {journal} {\bibinfo  {journal}
  {Astrophys. J.}\ }\textbf {\bibinfo {volume} {495}},\ \bibinfo {pages} {44}
  (\bibinfo {year} {1998})},\ \Eprint {http://arxiv.org/abs/astro-ph/9708018}
  {arXiv:astro-ph/9708018 [astro-ph]} \BibitemShut {NoStop}%
\bibitem [{\citenamefont {Croft}\ \emph {et~al.}(2002)\citenamefont {Croft},
  \citenamefont {Weinberg}, \citenamefont {Bolte}, \citenamefont {Burles},
  \citenamefont {Hernquist}, \citenamefont {Katz}, \citenamefont {Kirkman},\
  and\ \citenamefont {Tytler}}]{Croft:2000hs}%
  \BibitemOpen
  \bibfield  {author} {\bibinfo {author} {\bibfnamefont {R.~A.~C.}\
  \bibnamefont {Croft}}, \bibinfo {author} {\bibfnamefont {D.~H.}\ \bibnamefont
  {Weinberg}}, \bibinfo {author} {\bibfnamefont {M.}~\bibnamefont {Bolte}},
  \bibinfo {author} {\bibfnamefont {S.}~\bibnamefont {Burles}}, \bibinfo
  {author} {\bibfnamefont {L.}~\bibnamefont {Hernquist}}, \bibinfo {author}
  {\bibfnamefont {N.}~\bibnamefont {Katz}}, \bibinfo {author} {\bibfnamefont
  {D.}~\bibnamefont {Kirkman}}, \ and\ \bibinfo {author} {\bibfnamefont
  {D.}~\bibnamefont {Tytler}},\ }\href {\doibase 10.1086/344099} {\bibfield
  {journal} {\bibinfo  {journal} {Astrophys. J.}\ }\textbf {\bibinfo {volume}
  {581}},\ \bibinfo {pages} {20} (\bibinfo {year} {2002})},\ \Eprint
  {http://arxiv.org/abs/astro-ph/0012324} {arXiv:astro-ph/0012324 [astro-ph]}
  \BibitemShut {NoStop}%
  \bibitem [{\citenamefont {Vogelsberger}\ \emph {et~al.}(2016)\citenamefont
  {Vogelsberger}, \citenamefont {Zavala},\ \citenamefont {Cyr-Racine},\ 
  \citenamefont {Pfrommer},\ \citenamefont {Bringmann},\ and\ \citenamefont
  {Sigurdson}}]{Vogelsberger:2015gpr}%
  \BibitemOpen
  \bibfield  {author} {\bibinfo {author} {\bibfnamefont {M.}~\bibnamefont
  {Vogelsberger}}, \bibinfo {author} {\bibfnamefont {J.}~\bibnamefont
  {Zavala}}, \ \bibinfo {author} {\bibfnamefont {F.-Y.}~\bibnamefont
  {Cyr-Racine}}, \ \bibinfo {author} {\bibfnamefont {C.}~\bibnamefont
  {Pfrommer}}, \ \bibinfo {author} {\bibfnamefont {T.}~\bibnamefont
  {Bringmann}}, \ and\ \bibinfo {author} {\bibfnamefont {K.}~\bibnamefont
  {Sigurdson}},\ }\href {\doibase 10.1093/mnras/stw1076} {\bibfield
  {journal} {\bibinfo  {journal} {Mon. Not. Roy. Astron. Soc.}\ }\textbf
  {\bibinfo {volume} {460}},\ \bibinfo {pages} {1399} (\bibinfo {year}
  {2016})},\ \Eprint {http://arxiv.org/abs/1512.05349} {arXiv:1512.05349
  [astro-ph.CO]} \BibitemShut {NoStop}%
\bibitem [{\citenamefont {Bringmann}\ \emph {et~al.}(2016)\citenamefont
  {Bringmann}, \citenamefont {Ihle}, \citenamefont {Kersten},\ and\
  \citenamefont {Walia}}]{Bringmann:2016ilk}%
  \BibitemOpen
  \bibfield  {author} {\bibinfo {author} {\bibfnamefont {T.}~\bibnamefont
  {Bringmann}}, \bibinfo {author} {\bibfnamefont {H.~T.}\ \bibnamefont {Ihle}},
  \bibinfo {author} {\bibfnamefont {J.}~\bibnamefont {Kersten}}, \ and\
  \bibinfo {author} {\bibfnamefont {P.}~\bibnamefont {Walia}},\ }\href
  {\doibase 10.1103/PhysRevD.94.103529} {\bibfield  {journal} {\bibinfo
  {journal} {Phys. Rev.}\ }\textbf {\bibinfo {volume} {D94}},\ \bibinfo {pages}
  {103529} (\bibinfo {year} {2016})},\ \Eprint
  {http://arxiv.org/abs/1603.04884} {arXiv:1603.04884 [hep-ph]} \BibitemShut
  {NoStop}%
\bibitem [{\citenamefont {Ackermann}\ \emph {et~al.}(2015)\citenamefont
  {Ackermann} \emph {et~al.}}]{Ackermann:2015zua}%
  \BibitemOpen
  \bibfield  {author} {\bibinfo {author} {\bibfnamefont {M.}~\bibnamefont
  {Ackermann}} \emph {et~al.} (\bibinfo {collaboration} {Fermi-LAT}),\ }\href
  {\doibase 10.1103/PhysRevLett.115.231301} {\bibfield  {journal} {\bibinfo
  {journal} {Phys. Rev. Lett.}\ }\textbf {\bibinfo {volume} {115}},\ \bibinfo
  {pages} {231301} (\bibinfo {year} {2015})},\ \Eprint
  {http://arxiv.org/abs/1503.02641} {arXiv:1503.02641 [astro-ph.HE]}
  \BibitemShut {NoStop}%
\bibitem [{\citenamefont {Martinez}(2015)}]{Martinez:2013els}%
  \BibitemOpen
  \bibfield  {author} {\bibinfo {author} {\bibfnamefont {G.~D.}\ \bibnamefont
  {Martinez}},\ }\href {\doibase 10.1093/mnras/stv942} {\bibfield  {journal}
  {\bibinfo  {journal} {Mon. Not. Roy. Astron. Soc.}\ }\textbf {\bibinfo
  {volume} {451}},\ \bibinfo {pages} {2524} (\bibinfo {year} {2015})},\ \Eprint
  {http://arxiv.org/abs/1309.2641} {arXiv:1309.2641 [astro-ph.GA]} \BibitemShut
  {NoStop}%
\bibitem [{\citenamefont {Ackermann}\ \emph {et~al.}(2014)\citenamefont
  {Ackermann} \emph {et~al.}}]{Ackermann:2013yva}%
  \BibitemOpen
  \bibfield  {author} {\bibinfo {author} {\bibfnamefont {M.}~\bibnamefont
  {Ackermann}} \emph {et~al.} (\bibinfo {collaboration} {Fermi-LAT}),\ }\href
  {\doibase 10.1103/PhysRevD.89.042001} {\bibfield  {journal} {\bibinfo
  {journal} {Phys. Rev.}\ }\textbf {\bibinfo {volume} {D89}},\ \bibinfo {pages}
  {042001} (\bibinfo {year} {2014})},\ \Eprint {http://arxiv.org/abs/1310.0828}
  {arXiv:1310.0828 [astro-ph.HE]} \BibitemShut {NoStop}%
\bibitem [{\citenamefont {Accardo}\ \emph {et~al.}(2014)\citenamefont {Accardo}
  \emph {et~al.}}]{PhysRevLett.113.121101}%
  \BibitemOpen
  \bibfield  {author} {\bibinfo {author} {\bibfnamefont {L.}~\bibnamefont
  {Accardo}} \emph {et~al.} (\bibinfo {collaboration} {AMS Collaboration}),\
  }\href {\doibase 10.1103/PhysRevLett.113.121101} {\bibfield  {journal}
  {\bibinfo  {journal} {Phys. Rev. Lett.}\ }\textbf {\bibinfo {volume} {113}},\
  \bibinfo {pages} {121101} (\bibinfo {year} {2014})}\BibitemShut {NoStop}%
\bibitem [{\citenamefont {Aguilar}\ \emph {et~al.}(2014)\citenamefont {Aguilar}
  \emph {et~al.}}]{PhysRevLett.113.121102}%
  \BibitemOpen
  \bibfield  {author} {\bibinfo {author} {\bibfnamefont {M.}~\bibnamefont
  {Aguilar}} \emph {et~al.} (\bibinfo {collaboration} {AMS Collaboration}),\
  }\href {\doibase 10.1103/PhysRevLett.113.121102} {\bibfield  {journal}
  {\bibinfo  {journal} {Phys. Rev. Lett.}\ }\textbf {\bibinfo {volume} {113}},\
  \bibinfo {pages} {121102} (\bibinfo {year} {2014})}\BibitemShut {NoStop}%
\bibitem [{\citenamefont {Bergstrom}\ \emph {et~al.}(2013)\citenamefont
  {Bergstrom}, \citenamefont {Bringmann}, \citenamefont {Cholis}, \citenamefont
  {Hooper},\ and\ \citenamefont {Weniger}}]{Bergstrom:2013jra}%
  \BibitemOpen
  \bibfield  {author} {\bibinfo {author} {\bibfnamefont {L.}~\bibnamefont
  {Bergstrom}}, \bibinfo {author} {\bibfnamefont {T.}~\bibnamefont
  {Bringmann}}, \bibinfo {author} {\bibfnamefont {I.}~\bibnamefont {Cholis}},
  \bibinfo {author} {\bibfnamefont {D.}~\bibnamefont {Hooper}}, \ and\ \bibinfo
  {author} {\bibfnamefont {C.}~\bibnamefont {Weniger}},\ }\href {\doibase
  10.1103/PhysRevLett.111.171101} {\bibfield  {journal} {\bibinfo  {journal}
  {Phys. Rev. Lett.}\ }\textbf {\bibinfo {volume} {111}},\ \bibinfo {pages}
  {171101} (\bibinfo {year} {2013})},\ \Eprint {http://arxiv.org/abs/1306.3983}
  {arXiv:1306.3983 [astro-ph.HE]} \BibitemShut {NoStop}%
\bibitem [{\citenamefont {Elor}\ \emph {et~al.}(2016)\citenamefont {Elor},
  \citenamefont {Rodd}, \citenamefont {Slatyer},\ and\ \citenamefont
  {Xue}}]{Elor:2015bho}%
  \BibitemOpen
  \bibfield  {author} {\bibinfo {author} {\bibfnamefont {G.}~\bibnamefont
  {Elor}}, \bibinfo {author} {\bibfnamefont {N.~L.}\ \bibnamefont {Rodd}},
  \bibinfo {author} {\bibfnamefont {T.~R.}\ \bibnamefont {Slatyer}}, \ and\
  \bibinfo {author} {\bibfnamefont {W.}~\bibnamefont {Xue}},\ }\href {\doibase
  10.1088/1475-7516/2016/06/024} {\bibfield  {journal} {\bibinfo  {journal}
  {JCAP}\ }\textbf {\bibinfo {volume} {1606}},\ \bibinfo {pages} {024}
  (\bibinfo {year} {2016})},\ \Eprint {http://arxiv.org/abs/1511.08787}
  {arXiv:1511.08787 [hep-ph]} \BibitemShut {NoStop}%
\bibitem [{\citenamefont {Pospelov}\ and\ \citenamefont
  {Ritz}(2009)}]{Pospelov:2008jd}%
  \BibitemOpen
  \bibfield  {author} {\bibinfo {author} {\bibfnamefont {M.}~\bibnamefont
  {Pospelov}}\ and\ \bibinfo {author} {\bibfnamefont {A.}~\bibnamefont
  {Ritz}},\ }\href {\doibase 10.1016/j.physletb.2008.12.012} {\bibfield
  {journal} {\bibinfo  {journal} {Phys. Lett.}\ }\textbf {\bibinfo {volume}
  {B671}},\ \bibinfo {pages} {391} (\bibinfo {year} {2009})},\ \Eprint
  {http://arxiv.org/abs/0810.1502} {arXiv:0810.1502 [hep-ph]} \BibitemShut
  {NoStop}%
\bibitem [{\citenamefont {Feldman}\ \emph {et~al.}(2007)\citenamefont
  {Feldman}, \citenamefont {Kors},\ and\ \citenamefont
  {Nath}}]{Feldman:2006wd}%
  \BibitemOpen
  \bibfield  {author} {\bibinfo {author} {\bibfnamefont {D.}~\bibnamefont
  {Feldman}}, \bibinfo {author} {\bibfnamefont {B.}~\bibnamefont {Kors}}, \
  and\ \bibinfo {author} {\bibfnamefont {P.}~\bibnamefont {Nath}},\ }\href
  {\doibase 10.1103/PhysRevD.75.023503} {\bibfield  {journal} {\bibinfo
  {journal} {Phys. Rev.}\ }\textbf {\bibinfo {volume} {D75}},\ \bibinfo {pages}
  {023503} (\bibinfo {year} {2007})},\ \Eprint
  {http://arxiv.org/abs/hep-ph/0610133} {arXiv:hep-ph/0610133 [hep-ph]}
  \BibitemShut {NoStop}%
\bibitem [{\citenamefont {Frandsen}\ \emph {et~al.}(2011)\citenamefont
  {Frandsen}, \citenamefont {Kahlhoefer}, \citenamefont {Sarkar},\ and\
  \citenamefont {Schmidt-Hoberg}}]{Frandsen:2011cg}%
  \BibitemOpen
  \bibfield  {author} {\bibinfo {author} {\bibfnamefont {M.~T.}\ \bibnamefont
  {Frandsen}}, \bibinfo {author} {\bibfnamefont {F.}~\bibnamefont
  {Kahlhoefer}}, \bibinfo {author} {\bibfnamefont {S.}~\bibnamefont {Sarkar}},
  \ and\ \bibinfo {author} {\bibfnamefont {K.}~\bibnamefont {Schmidt-Hoberg}},\
  }\href {\doibase 10.1007/JHEP09(2011)128} {\bibfield  {journal} {\bibinfo
  {journal} {JHEP}\ }\textbf {\bibinfo {volume} {09}},\ \bibinfo {pages} {128}
  (\bibinfo {year} {2011})},\ \Eprint {http://arxiv.org/abs/1107.2118}
  {arXiv:1107.2118 [hep-ph]} \BibitemShut {NoStop}%
\bibitem [{\citenamefont {Foot}(2004)}]{Foot:2004pa}%
  \BibitemOpen
  \bibfield  {author} {\bibinfo {author} {\bibfnamefont {R.}~\bibnamefont
  {Foot}},\ }\href {\doibase 10.1142/S0218271804006449} {\bibfield  {journal}
  {\bibinfo  {journal} {Int. J. Mod. Phys.}\ }\textbf {\bibinfo {volume}
  {D13}},\ \bibinfo {pages} {2161} (\bibinfo {year} {2004})},\ \Eprint
  {http://arxiv.org/abs/astro-ph/0407623} {arXiv:astro-ph/0407623 [astro-ph]}
  \BibitemShut {NoStop}%
\bibitem [{\citenamefont {Redondo}\ and\ \citenamefont
  {Postma}(2009)}]{Redondo:2008ec}%
  \BibitemOpen
  \bibfield  {author} {\bibinfo {author} {\bibfnamefont {J.}~\bibnamefont
  {Redondo}}\ and\ \bibinfo {author} {\bibfnamefont {M.}~\bibnamefont
  {Postma}},\ }\href {\doibase 10.1088/1475-7516/2009/02/005} {\bibfield
  {journal} {\bibinfo  {journal} {JCAP}\ }\textbf {\bibinfo {volume} {0902}},\
  \bibinfo {pages} {005} (\bibinfo {year} {2009})},\ \Eprint
  {http://arxiv.org/abs/0811.0326} {arXiv:0811.0326 [hep-ph]} \BibitemShut
  {NoStop}%
\bibitem [{\citenamefont {Curtin}\ \emph {et~al.}(2015)\citenamefont {Curtin},
  \citenamefont {Essig}, \citenamefont {Gori},\ and\ \citenamefont
  {Shelton}}]{Curtin:2014cca}%
  \BibitemOpen
  \bibfield  {author} {\bibinfo {author} {\bibfnamefont {D.}~\bibnamefont
  {Curtin}}, \bibinfo {author} {\bibfnamefont {R.}~\bibnamefont {Essig}},
  \bibinfo {author} {\bibfnamefont {S.}~\bibnamefont {Gori}}, \ and\ \bibinfo
  {author} {\bibfnamefont {J.}~\bibnamefont {Shelton}},\ }\href {\doibase
  10.1007/JHEP02(2015)157} {\bibfield  {journal} {\bibinfo  {journal} {JHEP}\
  }\textbf {\bibinfo {volume} {02}},\ \bibinfo {pages} {157} (\bibinfo {year}
  {2015})},\ \Eprint {http://arxiv.org/abs/1412.0018} {arXiv:1412.0018
  [hep-ph]} \BibitemShut {NoStop}%
\bibitem [{\citenamefont {Chu}\ \emph {et~al.}(2012)\citenamefont {Chu},
  \citenamefont {Hambye},\ and\ \citenamefont {Tytgat}}]{Chu:2011be}%
  \BibitemOpen
  \bibfield  {author} {\bibinfo {author} {\bibfnamefont {X.}~\bibnamefont
  {Chu}}, \bibinfo {author} {\bibfnamefont {T.}~\bibnamefont {Hambye}}, \ and\
  \bibinfo {author} {\bibfnamefont {M.~H.~G.}\ \bibnamefont {Tytgat}},\ }\href
  {\doibase 10.1088/1475-7516/2012/05/034} {\bibfield  {journal} {\bibinfo
  {journal} {JCAP}\ }\textbf {\bibinfo {volume} {1205}},\ \bibinfo {pages}
  {034} (\bibinfo {year} {2012})},\ \Eprint {http://arxiv.org/abs/1112.0493}
  {arXiv:1112.0493 [hep-ph]} \BibitemShut {NoStop}%
\bibitem [{\citenamefont {Redondo}\ and\ \citenamefont
  {Raffelt}(2013)}]{Redondo:2013lna}%
  \BibitemOpen
  \bibfield  {author} {\bibinfo {author} {\bibfnamefont {J.}~\bibnamefont
  {Redondo}}\ and\ \bibinfo {author} {\bibfnamefont {G.}~\bibnamefont
  {Raffelt}},\ }\href {\doibase 10.1088/1475-7516/2013/08/034} {\bibfield
  {journal} {\bibinfo  {journal} {JCAP}\ }\textbf {\bibinfo {volume} {1308}},\
  \bibinfo {pages} {034} (\bibinfo {year} {2013})},\ \Eprint
  {http://arxiv.org/abs/1305.2920} {arXiv:1305.2920 [hep-ph]} \BibitemShut
  {NoStop}%
\bibitem [{\citenamefont {Garcia-Cely}\ and\ \citenamefont
  {Heeck}(2016)}]{Garcia-Cely:2016pse}%
  \BibitemOpen
  \bibfield  {author} {\bibinfo {author} {\bibfnamefont {C.}~\bibnamefont
  {Garcia-Cely}}\ and\ \bibinfo {author} {\bibfnamefont {J.}~\bibnamefont
  {Heeck}},\ }\href {\doibase 10.1088/1475-7516/2016/08/023} {\bibfield
  {journal} {\bibinfo  {journal} {JCAP}\ }\textbf {\bibinfo {volume} {1608}},\
  \bibinfo {pages} {023} (\bibinfo {year} {2016})},\ \Eprint
  {http://arxiv.org/abs/1605.08049} {arXiv:1605.08049 [hep-ph]} \BibitemShut
  {NoStop}%
\bibitem [{\citenamefont {Abbasi}\ \emph {et~al.}(2011)\citenamefont {Abbasi}
  \emph {et~al.}}]{IceCube:2011ae}%
  \BibitemOpen
  \bibfield  {author} {\bibinfo {author} {\bibfnamefont {R.}~\bibnamefont
  {Abbasi}} \emph {et~al.} (\bibinfo {collaboration} {IceCube}),\ }in\ \href
  {http://inspirehep.net/record/945595/files/arXiv:1111.2738.pdf} {\emph
  {\bibinfo {booktitle} {{Proceedings, 32nd International Cosmic Ray Conference
  (ICRC 2011): Beijing, China, August 11-18, 2011}}}}\ (\bibinfo {year}
  {2011})\ \Eprint {http://arxiv.org/abs/1111.2738} {arXiv:1111.2738
  [astro-ph.HE]} \BibitemShut {NoStop}%
\bibitem [{\citenamefont {Dasgupta}\ and\ \citenamefont
  {Laha}(2012)}]{Dasgupta:2012bd}%
  \BibitemOpen
  \bibfield  {author} {\bibinfo {author} {\bibfnamefont {B.}~\bibnamefont
  {Dasgupta}}\ and\ \bibinfo {author} {\bibfnamefont {R.}~\bibnamefont
  {Laha}},\ }\href {\doibase 10.1103/PhysRevD.86.093001} {\bibfield  {journal}
  {\bibinfo  {journal} {Phys. Rev.}\ }\textbf {\bibinfo {volume} {D86}},\
  \bibinfo {pages} {093001} (\bibinfo {year} {2012})},\ \Eprint
  {http://arxiv.org/abs/1206.1322} {arXiv:1206.1322 [hep-ph]} \BibitemShut
  {NoStop}%
\bibitem [{\citenamefont {Aartsen}\ \emph {et~al.}(2015)\citenamefont {Aartsen}
  \emph {et~al.}}]{Aartsen:2015xej}%
  \BibitemOpen
  \bibfield  {author} {\bibinfo {author} {\bibfnamefont {M.~G.}\ \bibnamefont
  {Aartsen}} \emph {et~al.} (\bibinfo {collaboration} {IceCube}),\ }\href
  {\doibase 10.1140/epjc/s10052-015-3713-1} {\bibfield  {journal} {\bibinfo
  {journal} {Eur. Phys. J.}\ }\textbf {\bibinfo {volume} {C75}},\ \bibinfo
  {pages} {492} (\bibinfo {year} {2015})},\ \Eprint
  {http://arxiv.org/abs/1505.07259} {arXiv:1505.07259 [astro-ph.HE]}
  \BibitemShut {NoStop}%
\bibitem [{\citenamefont {El~Aisati}\ \emph {et~al.}(2015)\citenamefont
  {El~Aisati}, \citenamefont {Gustafsson},\ and\ \citenamefont
  {Hambye}}]{Aisati:2015vma}%
  \BibitemOpen
  \bibfield  {author} {\bibinfo {author} {\bibfnamefont {C. E.}~\bibnamefont
  {Aisati}}, \bibinfo {author} {\bibfnamefont {M.}~\bibnamefont
  {Gustafsson}}, \ and\ \bibinfo {author} {\bibfnamefont {T.}~\bibnamefont
  {Hambye}},\ }\href {\doibase 10.1103/PhysRevD.92.123515} {\bibfield
  {journal} {\bibinfo  {journal} {Phys. Rev.}\ }\textbf {\bibinfo {volume}
  {D92}},\ \bibinfo {pages} {123515} (\bibinfo {year} {2015})},\ \Eprint
  {http://arxiv.org/abs/1506.02657} {arXiv:1506.02657 [hep-ph]} \BibitemShut
  {NoStop}%
\bibitem [{\citenamefont {Bringmann}\ \emph {et~al.}(2014)\citenamefont
  {Bringmann}, \citenamefont {Hasenkamp},\ and\ \citenamefont
  {Kersten}}]{Bringmann:2013vra}%
  \BibitemOpen
  \bibfield  {author} {\bibinfo {author} {\bibfnamefont {T.}~\bibnamefont
  {Bringmann}}, \bibinfo {author} {\bibfnamefont {J.}~\bibnamefont
  {Hasenkamp}}, \ and\ \bibinfo {author} {\bibfnamefont {J.}~\bibnamefont
  {Kersten}},\ }\href {\doibase 10.1088/1475-7516/2014/07/042} {\bibfield
  {journal} {\bibinfo  {journal} {JCAP}\ }\textbf {\bibinfo {volume} {1407}},\
  \bibinfo {pages} {042} (\bibinfo {year} {2014})},\ \Eprint
  {http://arxiv.org/abs/1312.4947} {arXiv:1312.4947 [hep-ph]} \BibitemShut
  {NoStop}%
\bibitem [{\citenamefont {Dasgupta}\ and\ \citenamefont
  {Kopp}(2014)}]{Dasgupta:2013zpn}%
  \BibitemOpen
  \bibfield  {author} {\bibinfo {author} {\bibfnamefont {B.}~\bibnamefont
  {Dasgupta}}\ and\ \bibinfo {author} {\bibfnamefont {J.}~\bibnamefont
  {Kopp}},\ }\href {\doibase 10.1103/PhysRevLett.112.031803} {\bibfield
  {journal} {\bibinfo  {journal} {Phys. Rev. Lett.}\ }\textbf {\bibinfo
  {volume} {112}},\ \bibinfo {pages} {031803} (\bibinfo {year} {2014})},\
  \Eprint {http://arxiv.org/abs/1310.6337} {arXiv:1310.6337 [hep-ph]}
  \BibitemShut {NoStop}%
\bibitem [{\citenamefont {Chu}\ and\ \citenamefont
  {Dasgupta}(2014)}]{Chu:2014lja}%
  \BibitemOpen
  \bibfield  {author} {\bibinfo {author} {\bibfnamefont {X.}~\bibnamefont
  {Chu}}\ and\ \bibinfo {author} {\bibfnamefont {B.}~\bibnamefont {Dasgupta}},\
  }\href {\doibase 10.1103/PhysRevLett.113.161301} {\bibfield  {journal}
  {\bibinfo  {journal} {Phys. Rev. Lett.}\ }\textbf {\bibinfo {volume} {113}},\
  \bibinfo {pages} {161301} (\bibinfo {year} {2014})},\ \Eprint
  {http://arxiv.org/abs/1404.6127} {arXiv:1404.6127 [hep-ph]} \BibitemShut
  {NoStop}%
\bibitem [{\citenamefont {Ko}\ and\ \citenamefont {Tang}(2014)}]{Ko:2014bka}%
  \BibitemOpen
  \bibfield  {author} {\bibinfo {author} {\bibfnamefont {P.}~\bibnamefont
  {Ko}}\ and\ \bibinfo {author} {\bibfnamefont {Y.}~\bibnamefont {Tang}},\
  }\href {\doibase 10.1016/j.physletb.2014.10.035} {\bibfield  {journal}
  {\bibinfo  {journal} {Phys. Lett.}\ }\textbf {\bibinfo {volume} {B739}},\
  \bibinfo {pages} {62} (\bibinfo {year} {2014})},\ \Eprint
  {http://arxiv.org/abs/1404.0236} {arXiv:1404.0236 [hep-ph]} \BibitemShut
  {NoStop}%
\bibitem [{\citenamefont {Binder}\ \emph {et~al.}(2016)\citenamefont {Binder},
  \citenamefont {Covi}, \citenamefont {Kamada}, \citenamefont {Murayama},
  \citenamefont {Takahashi},\ and\ \citenamefont {Yoshida}}]{Binder:2016pnr}%
  \BibitemOpen
  \bibfield  {author} {\bibinfo {author} {\bibfnamefont {T.}~\bibnamefont
  {Binder}}, \bibinfo {author} {\bibfnamefont {L.}~\bibnamefont {Covi}},
  \bibinfo {author} {\bibfnamefont {A.}~\bibnamefont {Kamada}}, \bibinfo
  {author} {\bibfnamefont {H.}~\bibnamefont {Murayama}}, \bibinfo {author}
  {\bibfnamefont {T.}~\bibnamefont {Takahashi}}, \ and\ \bibinfo {author}
  {\bibfnamefont {N.}~\bibnamefont {Yoshida}},\ }\href {\doibase
  10.1088/1475-7516/2016/11/043} {\bibfield  {journal} {\bibinfo  {journal}
  {JCAP}\ }\textbf {\bibinfo {volume} {1611}},\ \bibinfo {pages} {043}
  (\bibinfo {year} {2016})},\ \Eprint {http://arxiv.org/abs/1602.07624}
  {arXiv:1602.07624 [hep-ph]} \BibitemShut {NoStop}%
\bibitem [{\citenamefont {Feng}\ \emph {et~al.}(2008)\citenamefont {Feng},
  \citenamefont {Tu},\ and\ \citenamefont {Yu}}]{Feng:2008mu}%
  \BibitemOpen
  \bibfield  {author} {\bibinfo {author} {\bibfnamefont {J.~L.}\ \bibnamefont
  {Feng}}, \bibinfo {author} {\bibfnamefont {H.}~\bibnamefont {Tu}}, \ and\
  \bibinfo {author} {\bibfnamefont {H.-B.}\ \bibnamefont {Yu}},\ }\href
  {\doibase 10.1088/1475-7516/2008/10/043} {\bibfield  {journal} {\bibinfo
  {journal} {JCAP}\ }\textbf {\bibinfo {volume} {0810}},\ \bibinfo {pages}
  {043} (\bibinfo {year} {2008})},\ \Eprint {http://arxiv.org/abs/0808.2318}
  {arXiv:0808.2318 [hep-ph]} \BibitemShut {NoStop}%
\bibitem [{\citenamefont {Kaplinghat}\ \emph {et~al.}(2015)\citenamefont
  {Kaplinghat}, \citenamefont {Linden},\ and\ \citenamefont
  {Yu}}]{Kaplinghat:2015gha}%
  \BibitemOpen
  \bibfield  {author} {\bibinfo {author} {\bibfnamefont {M.}~\bibnamefont
  {Kaplinghat}}, \bibinfo {author} {\bibfnamefont {T.}~\bibnamefont {Linden}},
  \ and\ \bibinfo {author} {\bibfnamefont {H.-B.}\ \bibnamefont {Yu}},\ }\href
  {\doibase 10.1103/PhysRevLett.114.211303} {\bibfield  {journal} {\bibinfo
  {journal} {Phys. Rev. Lett.}\ }\textbf {\bibinfo {volume} {114}},\ \bibinfo
  {pages} {211303} (\bibinfo {year} {2015})},\ \Eprint
  {http://arxiv.org/abs/1501.03507} {arXiv:1501.03507 [hep-ph]} \BibitemShut
  {NoStop}%
\bibitem [{\citenamefont {deNiverville}\ \emph {et~al.}(2011)\citenamefont
  {deNiverville}, \citenamefont {Pospelov},\ and\ \citenamefont
  {Ritz}}]{deNiverville:2011it}%
  \BibitemOpen
  \bibfield  {author} {\bibinfo {author} {\bibfnamefont {P.}~\bibnamefont
  {deNiverville}}, \bibinfo {author} {\bibfnamefont {M.}~\bibnamefont
  {Pospelov}}, \ and\ \bibinfo {author} {\bibfnamefont {A.}~\bibnamefont
  {Ritz}},\ }\href {\doibase 10.1103/PhysRevD.84.075020} {\bibfield  {journal}
  {\bibinfo  {journal} {Phys. Rev.}\ }\textbf {\bibinfo {volume} {D84}},\
  \bibinfo {pages} {075020} (\bibinfo {year} {2011})},\ \Eprint
  {http://arxiv.org/abs/1107.4580} {arXiv:1107.4580 [hep-ph]} \BibitemShut
  {NoStop}%
\end{thebibliography}
\end{document}